\documentclass[usegraphicx,usenatbib,useAMS]{mn2e}
\usepackage{times}
\newcommand{\be}{\begin{equation}}
\newcommand{\ba}{\begin{eqnarray}}
\newcommand{\ee}{\end{equation}}
\newcommand{\ea}{\end{eqnarray}}

\newcommand{\url}{\tt}% 
\def\gtsima{$\; \buildrel > \over \sim \;$}
\def\ltsima{$\; \buildrel < \over \sim \;$}
\def\gsim{\lower.5ex\hbox{\gtsima}}
\def\lsim{\lower.5ex\hbox{\ltsima}}
\def\simgt{\lower.5ex\hbox{\gtsima}}
\def\simlt{\lower.5ex\hbox{\ltsima}}
\def\simpr{\lower.5ex\hbox{\prosima}}

\def\simless{\mathbin{\lower 3pt\hbox
   {$\rlap{\raise 5pt\hbox{$\char'074$}}\mathchar''7218$}}}  % < or of order
\def\simgreat{\mathbin{\lower 3pt\hbox
   {$\rlap{\raise 5pt\hbox{$\char'076$}}\mathchar''7218$}}}  % > or of order
\def\apj{ApJ}
\def\apjs{ApJS}
\def\apjl{ApJL}
\def\aap{A\&A}
\def\aj{AJ}
\def\mnras{MNRAS}

\begin{document}

\title[Self-Regulated Reionization]{Self-Regulated Reionization}
\author[I.~T.~Iliev, et al.]
{Ilian T. Iliev$^{1}$\thanks{e-mail: iliev@cita.utoronto.ca},
Garrelt Mellema$^{2}$,
Paul~R.~Shapiro$^3$,
Ue-Li~Pen$^1$
\\
$^1$ Canadian Institute for Theoretical Astrophysics, University
  of Toronto, 60 St. George Street, Toronto, ON M5S 3H8, Canada\\
$^2$ Stockholm Observatory, AlbaNova University Center,
Stockholm University, SE-106 91 Stockholm, Sweden\\
$^3$ Department of Astronomy, University of Texas,
  Austin, TX 78712-1083
}
%% \author{Garrelt~Mellema$^{1,2}$, Ilian~T.~Iliev$^{3}$, Ue-Li~Pen$^3$, 
%% Paul~R.~Shapiro$^4$} 
%% \altaffiltext{1}{ASTRON, P.O. Box 1, NL-7990 AA Dwingeloo, The Netherlands}
%% \altaffiltext{2}{Sterrewacht Leiden, P.O. Box 9513, NL-2300 RA Leiden, The 
%% Netherlands}
%% \altaffiltext{3}{Canadian Institute for Theoretical Astrophysics, University
%%   of Toronto, 60 St. George Street, Toronto, ON M5S 3H8, Canada}
%% \altaffiltext{4}{Department of Astronomy, University of Texas, Austin, TX 
%% 78712-1083}
\date{\today, MNRAS accepted}
\pubyear{2006} \volume{000} \pagerange{1}
\twocolumn
\maketitle\label{firstpage}

\begin{abstract}
Recently, we have presented the first, truly large-scale radiative transfer 
simulations of inhomogeneous cosmic reionization which resolve
all the possible halo sources down to the dwarf galaxy mass range, 
$M\gsim10^9M_\odot$, in a comoving volume $(100\,h^{-1}\rm Mpc)^3$. This is 
large enough to sample the global mean history, geometry and statistical 
properties of reionization fairly and accurately for the first time. 
Here we present new simulations which extend the source halo mass range
downward to $10^8M_\odot$, to capture the full range of halo masses
thought to be primarily responsible for reionization by their star formation
following atomic hydrogen radiative cooling and gravitational collapse.
Haloes below about $10^9M_\odot$, however, are subject to Jeans-mass filtering 
in the ionized regions, which suppresses their baryonic content and their 
ability to release ionizing radiation.  By including these smaller-mass haloes
but accounting for their suppression, too, we find that reionization is
``self-regulating,'' as follows.  As the mean ionized fraction rises, so does
the fraction of the volume within which suppression occurs.  Hence, the
degree of suppression is related to the mean ionized fraction.  Since low-mass
haloes with high efficiency (i.e. high emissivity) achieve a given mean
ionized fraction earlier than do those with low efficiency, Jeans-mass
filtering compensates for the difference in the emissivity of the suppressible
haloes in these two cases. As a result, in the presence of 
lower-mass source haloes, reionization begins earlier, but the later stages
of reionization and the time of overlap are dictated by the efficiency of the 
higher-mass haloes, independent of the efficiency of the suppressible,
lower-mass haloes. Hence, while the lower-mass haloes do not alter the overlap 
redshift, $z_{\rm ov}$, they serve to boost the electron-scattering optical
depth of the universe, $\tau_{\rm es}$. This may explain why observations of
quasar absorption spectra at high redshift  
find that reionization ended late ($z_{\rm ov} < 7$), while CMB polarization 
measurements report a large enough $\tau_{\rm es}$ that reionization must have 
begun much earlier ($z > 11$).  We present results for the $\Lambda$CDM
universe with cosmological parameters from both 1-year and 3-year data
releases of WMAP.   
Reionization histories consistent with current constraints on $z_{\rm ov}$ and 
$\tau_{\rm es}$ are shown to be achievable with standard stellar sources in 
haloes above $10^8M_\odot$.  Neither minihalos nor exotic sources are required,
and the phenomenon of ``double reionization'' previously suggested does not occur.
\end{abstract}

\begin{keywords}
cosmology: theory --- radiative transfer ---
intergalactic medium --- large-scale structure of universe ---
galaxies: formation --- radio lines: galaxies
\end{keywords}

\section{Introduction}
The inhomogeneous reionization of the intergalactic medium (IGM) at high
redshift proceeded by the propagation of ionization fronts (I-fronts) outward
from the early galaxies that formed sources of ionizing radiation, like stars
and mini-quasars. This process continued until the H~II regions bounded by
these I-fronts grew to overlap. \citet{1987ApJ...321L.107S} showed that the
I-fronts during reionization were generally weak, R-type, which move
supersonically relative to both the neutral gas ahead of and the ionized gas
behind them, outracing the hydrodynamic response of the IGM to the radiation.
This property makes it possible to simulate inhomogeneous reionization in a
cosmological universe like $\Lambda$CDM by performing a radiative transfer
calculation on a pre-computed, 3-D, cosmological density field generated by
either a pure N-body or a gas and N-body dynamics simulation of large-scale 
structure formation. In
this approximation -- the ``static limit'' -- the dark matter and baryonic gas
evolve just as they would have without reionization, with no back-reaction
from the gas pressure forces associated with photoheating in an inhomogeneous
density field. These structure formation simulations also yield
the mass and location of the galactic halos which are the sites of star
formation and, hence, of ionizing photon release. Attempts to simulate
reionization in a CDM universe this way include \citet{2000MNRAS.314..611C,
2001MNRAS.321..593N,2002ApJ...572..695R,2003MNRAS.343.1101C,
2003MNRAS.344..607S,2006MNRAS.369.1625I,21cmreionpaper}. Of these, the simulations
described in \citet{2006MNRAS.369.1625I} and \citet{21cmreionpaper} represent the 
first truly large-scale radiative transfer simulations of reionization, in a 
comoving volume of $(100\,\rm h^{-1}Mpc)^3$, which resolve all the individual halo
sources down to the dwarf galaxy range, above about a billion solar masses.
This represents a more than two orders of magnitude volume increase over
previous work, as is required in order to make statistically meaningful
predictions of observable consequences and features of reionization. To
accomplish this, a new and efficient radiative transfer method, called
C$^2$-Ray (Conservative, Causal Ray-Tracing) was developed
\citep{methodpaper}. It was then coupled to the results of very large N-body
simulations of the $\Lambda$CDM universe involving $1624^3=4.28$ billion
particles on a cubic lattice of $3248^3$ cells by the parallel PMFAST method
\citep{2005NewA...10..393M}. 

The back-reaction of the IGM and of the halos which collapse out of it to make
sources cannot be entirely neglected, however. These effects become important
at small scales, in particular. As first described in
\citet[][henceforth, SGB94]{1994ApJ...427...25S}, the heating of the IGM during
reionization introduces pressure forces which oppose the growth of linear
perturbations in the baryonic component by gravitational instability --
sometimes referred to as ``Jeans-mass filtering'' -- resulting in a 
negative feedback on the rate of collapse of new baryons out of the IGM into
dark matter halos. Since small-mass halos form first and merge to make
larger-mass halos later during hierarchical structure formation in the
$\Lambda$CDM universe, this feedback effect (which prevents the smaller-mass 
halo sources from forming inside H~II regions) acts as a ``self-regulator'' on 
reionization, as follows. The
more sources that form, the higher the mean ionized fraction of the universe,
but the higher the fraction of the IGM which is ionized, the more of these
small-mass halos are suppressed. This slows the rapid exponential rise of the 
source population and, with it, the rise of the ionized fraction. Eventually,
when halos massive enough to overcome this feedback effect and capture their
fair share of baryons even inside H~II regions become common enough, they are 
able to finish reionization without being inhibited by radiative feedback. This
phenomenon of Jeans-mass filtering was subsequently revisited and confirmed
by comparison with simulations by
\citet{1998MNRAS.296...44G} and \citet{2000ApJ...542..535G}, see also 
\citet{1997ApJ...479..523B}. Estimates of the 
minimum halo mass scale which is sufficient to overcome the effects of uniform 
photoionization and accrete or retain enough baryons to make star formation 
possible have
also been refined over the years, by 1-D and 3-D gasdynamical simulations
\citep[e.g.][]{1992MNRAS.256P..43E,1996ApJ...465..608T,1997ApJ...478...13N,
2004ApJ...601..666D}. Roughly speaking, for halos during the epoch of
reionization (EOR) at redshifts $z>6$, the minimum halo mass required to avoid
suppression is about a billion solar masses. 

This mass scale corresponds roughly to the minimum mass of the source halos
included in the large-scale reionization simulations described in
\citet{2006MNRAS.369.1625I} and \citet{21cmreionpaper} for comoving simulation 
volumes of $(100\,\rm h^{-1}Mpc)^3$, those of
\citet{2003MNRAS.343.1101C,2003MNRAS.344L...7C,MH_sim}, for simulation volumes
$(20\,\rm h^{-1}Mpc)^3$ and those of \citet{2006astro.ph..4177Z} for a volume
of $(65\,\rm h^{-1}Mpc)^3$. For these simulations the Jeans-mass filtering is
not taken into account, since the mass range of the source halos they resolved
is above the range which is suppressed by reionization. 

The minimum mass of the halos responsible for reionization may have 
been smaller than this, however. The star-forming abilities of halos are 
generally thought to be different for halos with virial temperatures 
$T_{\rm vir}$
above and below $10^4$~K, respectively. Halos with $T_{\rm vir}<10^4$~K 
(roughly $10^4\simlt M/M_\odot\simlt 10^8$) called ``minihaloes'' only 
form stars if they can form a trace of $H_2$ molecules sufficient to
cool the gas well below $T_{\rm vir}$ by collisional excitation of 
rotational-vibrational lines. Since minihaloes are the earliest halos to 
form in the CDM universe, they are expected to be the first sites of star 
formation. Minihaloes are highly vulnerable to radiative feedback from 
these very stars, however. Simulations currently suggest that these first 
stars were massive and hot and, hence, were strong emitters of ionizing 
and dissociating UV radiation \citep{2002Sci...295...93A,2002ApJ...564...23B}. 
A single star was then sufficient to expel the gas from the minihalo that 
made it, by photoionizing the gas \citep{2004ApJ...610...14W,2004ApJ...613..631K,
2006ApJ...639..621A}. The UV background from these stars in the Lyman-Werner bands 
is likely to have dissociated the $H_2$ molecules in minihaloes which had 
not yet formed stars, long before they released enough ionizing photons to
reionize the universe \citep{2000ApJ...534...11H}. This suggests that minihaloes 
were not 
the primary source of reionization, although this conclusion remains highly 
uncertain. Those minihaloes which were ``sterilized'' in this way against 
their own star formation would have trapped the intergalactic I-fronts which
encountered them during reionization, transforming these I-fronts from R-type
to D-type, and expelling their gas in a photoevaporative wind 
\citep{2004MNRAS.348..753S,2005MNRAS...361..405I}.
This process might have affected the progress of reionization by consuming
additional ionizing photons 
\citep{2005MNRAS...361..405I,2005ApJ...624..491I,MH_sim}. 
The effect of 
minihaloes as sinks of ionizing photons tends to be degenerate, in fact, 
with the unknown efficiency for release of ionizing photons by higher-mass 
sources, as a result of the tendency for minihaloes to cluster around them  
\citep{2005ApJ...624..491I}.

By contrast with the minihaloes, halos with $T_{\rm vir}>10^4$~K 
($M \simgt10^8M_\odot$) were less vulnerable to the suppression of their 
star formation. These halos were able to cool radiatively and collapse
by collisional excitation of the atomic hydrogen Ly$\alpha$ line, leading
to star formation even in the presence of the rising UV dissociation 
background from other stars. As described above, however, these halos 
were nevertheless subject to Jeans-mass filtering inside H~II regions.

Halos in the intermediate mass range ($10^8\simlt M/M_\odot \simlt 10^9$),
therefore, can be important sources of additional ionizing photons not 
explicitly accounted for in our previous large-scale simulations of 
reionization. These sources would have been suppressed inside H~II regions,
however, so their overall impact remains to be determined. Semi-analytical
studies suggest that the inclusion of halos subject to Jeans-mass filtering
has a significant effect on the mean reionization history 
\citep{1994ApJ...427...25S,1995ASPC...80...55S,2000ApJ...534..507C,
2003ApJ...595....1H,2003ApJ...586..693W,2004ApJ...610....1O,
2005ApJ...634....1F}.
This may help explain why the CMB polarization experiments indicate that 
reionization was well advanced by $z\simgt11$ \citep{2006astro.ph..3449S}, 
while quasar absorption spectra suggest that reionization ended later, at 
$z\sim6.5$ \citep[e.g.][]{2002AJ....123.1247F,2003AJ.126..1W}, implying that 
the EOR was extended in time. 

Some smaller-scale reionization simulations have been performed that couple
gas and N-body dynamics to radiative transfer, which, in principle, takes 
account of the feedback effect of Jeans-mass filtering \citep{2002ApJ...575...49R}. 
However, these simulation volumes are too small to describe the global 
reionization history or its geometry and statistical properties. As a result,
they failed to anticipate the extended nature of reionization which 
self-regulation by Jeans-mass filtering makes
possible\footnote{Recently, small-scale reionization simulations like these
  were also used as the basis for a phenomenological ``subgrid model'' in a
  large-scale reionization simulation which was too coarse-grained to resolve
  the small-scale structure and source halos directly \citep{2005astro.ph.11627K}.}
That will be the subject of the current paper.    

In this paper we assume a flat ($\Omega_k=0$) $\Lambda$CDM cosmology with 
parameters 
($\Omega_m,\Omega_\Lambda,\Omega_b,h,\sigma_8,n)=(0.27,0.73,0.044,0.7,0.9,1)$
\citep{2003ApJS..148..175S}, hereafter WMAP1, where 
$\Omega_m$, $\Omega_\Lambda$, and $\Omega_b$ are the total matter, vacuum, 
and baryonic densities in units of the critical density, $\sigma_8$ is the 
rms density fluctuations extrapolated to the present on the scale of 
$8 h^{-1}{\rm Mpc}$ according to the linear perturbation theory, and $n$ is 
the index of the primordial power spectrum of density fluctuations. The new, 
3-year WMAP data yielded somewhat different parameters
\citep{2006astro.ph..3449S}, ($\Omega_m,\Omega_\Lambda,\Omega_b,h,\sigma_8,n)=
(0.24,0.76,0.042,0.73,0.74,0.95)$, hereafter WMAP3. 

The differences between these WMAP1 and WMAP3 parameters can have a
significant impact on the progress of reionization
\citep{2006ApJ...644L.101A}. Structure formation is delayed in 
the WMAP3 universe relative to the WMAP1 universe, especially on the small
scales relevant to the formation of reionization source halos at high
redshift, so the epoch of reionization is shifted to lower redshifts. 
In particular, if source halos of a given mass are assumed to have released 
ionizing photons with the same efficiency in either case, then reionization 
for WMAP3 is predicted to have occurred at $(1+z)$-values which are roughly 
1.4 times smaller than for WMAP1. As such, the predicted electron-scattering 
optical depth of the IGM accumulated since the beginning of the EOR would have
been smaller for WMAP3 than for WMAP1 by a factor of $1.4^{3/2}\sim1.7$, just
as the observations of large-angle fluctuations in the CMB polarization
require. This means that the ionizing efficiency per collapsed baryon required
to make reionization early enough to explain the value of $\tau_{\rm es}$
reported for WMAP1 and WMAP3 are nearly the same, despite the fact that
$\tau_{\rm es}$ is smaller for WMAP3 than for WMAP1.

     In Section 2, we describe our simulations, the method, input
parameters and cases.  We motivate these simulations and anticipate some
of the trends by a simple analytical toy model for the mean ionization
history in Section 3.  Simulation results are presented in Section 4,
with conclusions in Section 5.  Appendix A describes how we can use the
simulation results to improve upon the toy model introduced in Section 3.

\section{The Simulations}
Our basic methodology was described in detail in \citet{2006MNRAS.369.1625I}
(hereafter Paper I) and \citet{21cmreionpaper} (hereafter Paper II).
Hence, here we will outline the main simulation parameters and 
concentrate on the simulations which have not been presented before and 
the new features we introduce. 

We use very high resolution N-body simulations to derive halo catalogues, 
which include the detailed halo properties, as well as their positions and
velocities at a number of time-slices (between 50 and 100 per simulation) 
and the corresponding gas density fields. All the halos found in the 
simulation volume at a given redshift are assumed to be sources of ionizing 
radiation (unless they are suppressed by Jeans mass filtering, see 
below). This results typically in tens to hundreds of thousands of sources, 
depending on the case. We use a simple recipe to assign a photon emissivity 
to each source, by assuming a constant mass-to-light ratio. We follow the 
time-dependent propagation of the ionization fronts produced by all sources in
the simulation volume using our detailed radiative transfer and
non-equilibrium chemistry code $C^2$-Ray~\citep{methodpaper}, which has been
extensively tested against available analytical solutions~\citep{methodpaper}
and a number of other cosmological radiative transfer codes
\citep{comparison1}. 

The underlying N-body simulations have a spatial grid of $3248^3$ cells and
follow the evolution of $1624^3=4.3$ billion particles using the particle-mesh
code PMFAST \citep{2005NewA...10..393M}. The number of resolved halos for
the simulations with WMAP1 cosmology parameters are $\sim4\times10^5$ 
($\sim8\times10^5$) halos at $z=8$ ($z=6$) in the $(100\,\rm h^{-1}~Mpc)^3$ 
volume, and $\sim7\times10^5$ ($\sim9\times10^5$) halos at $z=8$ ($z=6$) 
 in the $(35\,\rm h^{-1}~Mpc)^3$ volume. The corresponding numbers for the 
WMAP3 simulations are $\sim7.5\times10^4$ ($\sim3\times10^5$) halos at $z=8$ 
($z=6$) in the $(100\,\rm h^{-1}~Mpc)^3$ volume and $\sim2\times10^5$ 
($\sim3\times10^5$) halos at $z=8$ ($z=6$) in the $(35\,\rm h^{-1}~Mpc)^3$ 
volume.

Simulating the transfer of ionizing radiation 
with the same grid resolution as the underlying N-body is still not feasible 
on current computers. We therefore re-grid the data to lower resolution, with
either $203^3$ or $406^3$ cells, for the radiative transfer simulations. We
combine sources which fall into the same coarse cell, which reduces slightly 
the number of sources to be considered compared to the total number of halos.

\subsection{Source suppression by Jeans-mass filtering}

The process of photoionization also heats the gas to temperatures above
$10^4$~K. The exact value of the temperature reached varies and generally 
depends on the local level of the ionizing flux and its spectrum
\citep[see][for detailed numerical calculations]{2004MNRAS.348..753S}. Typical
values are $T_{\rm IGM}=10,000-20,000$~K, but it could be as high as
$\sim40,000$~K for hot (Pop.~III) black-body spectrum. However, as was
mentioned above, the hydrogen line cooling is highly efficient for $T>10^4$~K,
particularly at high redshifts, where the gas is denser on average, which
would typically bring its temperature down to $T_{\rm IGM}\sim10^4$~K, and
possibly somewhat below that due to the adiabatic cooling from the expansion
of the universe. 

This increase of the IGM temperature caused by its photoheating results in a
corresponding increase in the Jeans mass. The adiabatic IGM temperature at
$z\simlt130$, after Compton scattering ceases to couple $T_{\rm IGM}$ to
$T_{\rm CMB}$ as it does at earlier times, is well-approximated by
\be
T_{\rm IGM,0}=26{\rm\,mK}(1+z)^2
\ee   
\citep{1985MNRAS.214..137C,2003MNRAS.340..210G}. 
In linear theory, the instantaneous cosmological Jeans mass of the neutral IGM
in the absence of heating is then given by
\be
M_{J,0}=5340\,M_\odot\left(\frac{\Omega_0h^2}{0.15}\right)^{-1/2}
\left(\frac{\Omega_bh^2}{0.0223}\right)^{-3/5}\left(\frac{1+z}{10}\right)^{3/2},
\label{Jeans}
\ee
\citep[e.g.][]{1994ApJ...427...25S,2002ApJ...572L.123I}, and increases with 
temperature as $M_J\propto (T_{\rm IGM}/\mu)^{3/2}$, where $\mu$ is the mean 
molecular weight. Using this simple scaling with temperature we obtain 
\ba
M_J(T)\!\!\!&=&
\!\!\!M_{J,0}\left(\frac{T_{\rm IGM}}{T_{\rm
      IGM,0}}\right)^{3/2}\left(\frac{\mu}{\mu_0}\right)^{-3/2} 
\nonumber\\
&=&3.8\times10^9M_\odot\left(\frac{T_{\rm IGM}}{10^4\rm K}\right)^{3/2}
\left(\frac{\Omega_0h^2}{0.15}\right)^{-1/2}\\
&& \times \left(\frac{\Omega_bh^2}{0.0223}\right)^{-3/5}
   \left(\frac{1+z}{10}\right)^{3/2}\nonumber,
\label{MJ}
\ea
for the ionized IGM.

The actual filter mass differs somewhat from this instantaneous Jeans mass
since the mass scale on which baryons succeed in collapsing out of the IGM along
with the dark matter must be determined, even in linear theory, by integrating the
differential equation for perturbation growth over time for the evolving IGM 
\citep{1994ApJ...427...25S,1998MNRAS.296...44G,2000ApJ...542..535G}.
In reality, determining the minimum mass necessary for a halo collapsing inside 
an ionized and heated region to acquire its fair share of baryons which 
subsequently cool further to form stars is even more complicated. It depends on
the detailed, non-linear, gas dynamics of the process and on radiative cooling. 
There is no single mass above
which a collapsing halo retains all its gas, and below which the gas does not
collapse with the dark matter. Instead, simulations show that the cooled gas
fraction in halos decreases gradually with decreasing halo mass
\citep{1992MNRAS.256P..43E,1996ApJ...465..608T,1997ApJ...478...13N,
2004ApJ...601..666D}. 
The typical halo sizes at which this transition occurs as derived by these 
different studies also vary. \citet{1996ApJ...465..608T} found that photoionization
suppresses star formation in halos with circular velocities below
$\sim30\rm\,km\,s^{-1}$, and decreases the cooled gas mass fraction in larger
halos, with circular velocities up to $\sim50\rm\,km\,s^{-1}$.
The halo circular velocity is related to its virial temperature as follows
\be
T_{\rm vir}=
3.072\times10^4\rm\,K\left(\frac{\mu}{0.59}\right)\left(\frac{v_c}{30\rm
    km\,s^{-1}}\right)^2, 
\ee
\citep[e.g.][]{2001MNRAS.325..468I}, where $\mu=0.59$ is the mean molecular 
weight for ionized gas, and the virial temperature is related to the halo mass by
\be
T_{\rm vir}=
4.17\times10^4\rm\,K\left(\frac{\mu}{0.59}\right)
\left(\frac{M}{10^9M_\odot}\right)^{2/3}
\left(\frac{\Omega_0h^2}{0.15}\right)^{1/3}\left(\frac{1+z}{10}\right).  
\ee
\citet{1997ApJ...478...13N} found that the cooled gas fraction is
affected by photoionization even in larger galaxies, with circular velocities
up to $\sim100-200\rm\,km\,s^{-1}$. On the other hand,
\citet{2004ApJ...601..666D} recently showed, using the same method as
\citet{1996ApJ...465..608T}, that at high redshifts the suppression is not as 
effective, and somewhat smaller galaxies can still retain some cooled gas. For
simplicity, we assume that star formation is suppressed in halos with masses
below $10^9M_\odot$ and not suppressed in larger halos, in rough agreement
with the linear Jeans mass estimate for $10^4$~K gas and the above dynamical
studies. 

\subsection{Source efficiencies and Pop. III to Pop. II transition}
\label{eff_sect}

We model the sources by assuming a constant mass-to-light ratio. Each 
halo found in the simulation volume at a given time, which is not suppressed
by Jeans mass filtering is a source. For a source with halo mass $M$ and 
lifetime $t_s$ we assign ionizing photon emissivity according to
\be
\dot{N}_\gamma=f_\gamma\frac{M\Omega_b}{\mu m_pt_s\Omega_0},
\ee
where the proportionality coefficient $f_{\gamma}$ reflects the ionizing 
photon production efficiency of the stars per stellar atom, $N_i$, the star 
formation efficiency, $f_*$, and the escape fraction, $f_{\rm esc}$:
\be
f_\gamma=f_*f_{\rm esc}N_i.
\ee
\citep[e.g.][]{2003ApJ...595....1H}.
All these quantities are still quite uncertain, especially at high redshift,
see e.g. \citet{2005ApJ...624..491I} for discussion. Recent theoretical
studies have indicated that the first, metal-free stars (Pop. III) might have 
been quite massive \citep{2002ApJ...564...23B,2000ApJ...540...39A}. 
Massive stars are more efficient producers of ionizing photons, emitting
up to $N_i\sim10^5$ ionizing photons per stellar atom
\citep{2001ApJ...552..464B,2002A&A...382...28S,2003ApJ...594L...1V}. 
Integrating over a top-heavy IMF for Pop. III stars leads to estimates of 
$N_i\sim25,000-90,000$ \citep{2002A&A...382...28S}. The Salpeter IMF for
Pop. II stars gives
$N_i=3,000-10,000$ \citep{1999ApJS..123....3L}. The values of $f_*$ and 
$f_{\rm esc}$ are even less certain, ranging from $\sim0.01$ to $\sim1$ for 
each of these quantities. There are also indications that the photon escape
fraction is mass-dependent and significantly higher for small galaxies at
high redshift than for large galaxies observed at later times
\citep{2004ApJ...613..631K,2006ApJ...639..621A}. Thus, many of the
currently viable reionization scenarios involve an early population of small
sources with high ionizing efficiency, which eventually evolve into the
population of less efficient emitters we see at later times. In this work we 
adopt $f_\gamma=2000$ (corresponding to e.g. $N_i=50,000$, $f_*=0.2$ and
$f_{\rm esc}=0.2$, i.e. top-heavy IMF and relatively efficient star formation
and photon escape) for modelling the high-efficiency emitters and
$f_\gamma=250$ (corresponding to e.g. $N_i=25,000$, $f_*=0.1$ and 
$f_{\rm esc}=0.1$, or $N_i=6,000$, $f_*=0.2$ and $f_{\rm esc}=0.21$,
i.e. either moderately top-heavy IMF and moderate efficiencies, or Salpeter
IMF and relatively high, but not unreasonable, efficiencies). 

The detailed mechanisms of this possible transition from high to low
efficiency emitters, and even the question of whether it actually occurred are
still unclear. We 
therefore have simulated cases both with and without such a transition. In the
latter case, all sources, both large and small, have the same ionizing photon
production efficiency. The possible physical mechanisms which result in
decreasing the ionizing efficiency are quite varied and complex, including
e.g.  production, expulsion and mixing of metals, which modifies the stellar 
IMF, and thus $N_i$, and the increase of the mean halo mass over time as a
consequence of hierarchical structure formation, which is expected to decrease
the escape
fraction. Modelling and studying the detailed features and mechanisms of this
transition is not the aim of this paper. Instead of trying to model these
complex processes and the many related uncertainties, we adopt a simple model
which should nonetheless capture the main features of the efficiency
transition, as follows. For the simulations with varying ionizing efficiency,
we assign the high value of $f_\gamma$ to the sources smaller than some
characteristic mass, and the low efficiency to the larger halos. Since the 
CDM structure formation proceeds hierarchically, the small halos form first, 
and gradually merge up to form ever larger halos. This process of merging
is presumed to lead to a gradual increase in the mean metallicity of the halo 
gas, and decrease of the escape fractions. Physically, it is expected that
this transition is inhomogeneous in space and extended in time
\citep[e.g.][]{2005ApJ...634....1F} and proceeded faster in the high-density 
peaks, which are the first to form halos and are the first sites of vigorous 
merging. We assume the characteristic 
halo mass at which the efficiency transition occurs to be $10^9M_\odot$. For 
computational simplicity, we chose it to be the same one as the Jeans
filtering mass since this procedure yields only two types of sources (small,
efficient and suppressible and large, less efficient and unsuppressible),
rather than multiple source types. This assumption does not affect any of our 
main qualitative conclusions, though the exact value of the characteristic
mass may affect some of the detailed quantitative estimates. If this
transition mass were somewhat lower, e.g. $5\times10^8 M_\odot$, rather than 
$10^9 M_\odot$, then the largest of the suppressed halos (between $5\times10^8
M_\odot$ and $10^9 M_\odot$) would have Pop. II efficiency, while 
the smaller ones would have Pop. III efficiency. This would simply yield a
case somewhere in-between our cases of high and low Pop. III efficiencies and 
our main conclusions would hold, although some quantitative numbers would 
change.

\begin{table*}
\caption{Simulation parameters and global reionization history results
for simulations with WMAP1 cosmology parameters. 
Box sizes are in [$\,h^{-1}$Mpc].}
\label{summary}
\begin{center}
\begin{tabular}{@{}llllllllllll}\hline
                         & f2000 & f2000\_406& f250   & f2000C & f250C  & f2000\_250 & f2000\_250S & f2000\_250S\_406& f250\_250S& f2000C\_250S& f250C\_250S \\[2mm]
\hline\\
mesh                     & $203^3$ & $406^3$ & $203^3$& $203^3$& $203^3$& $203^3$    & $203^3$     & $406^3$         &     $203^3$ & $203^3$   & $203^3$     \\[2mm]
box size                 & 100     & 100     & 100    & 100    & 100    & 35         & 35          & 35              & 35          & 35        & 35          \\[2mm]
$(f_\gamma)_{\rm large}$ & 2000    & 2000    & 250    & 2000   & 250    & 250        & 250         & 250             & 250         & 250       & 250         \\[2mm]
$(f_\gamma)_{\rm small}$ & -       & -       & -      & -      & -      & 2000       & 2000        & 2000            & 250         & 2000      & 250         \\[2mm]
$C_{\rm subgrid}$        & 1       & 1       & 1      & $C(z)$ & $C(z)$ & 1          & 1           & 1               & 1           & $C(z)$    & $C(z)$      \\[2mm]
Jeans supp.              & -       & -       & -      & -      & -      & no         & yes         & yes             & yes         & yes       & yes         \\[2mm]
\hline\\ 
$z_{50\%}$               & 13.6 & 13.5       & 11.7   & 12.6   & 11     & 16.2       & 14.5        & 14.9            &12.6         &13.8       & 11.6        \\[2mm] 
$z_{\rm overlap}$        & 11.3 & $\sim 11$  & 9.3    & 10.2   &$8.2$   & 13.5       & 10.4        & 10.4            &9.9          &9.1        & 8.4         \\[2mm] 
%$\tau_{\rm es}$          & 0.145& $\sim 0.14$& 0.121  & 0.135  &0.107   & 0.197      & 0.167       & 0.169           &0.138        &0.151      & 0.122       \\[2mm]
$\tau_{\rm es}$          & 0.130& $\sim 0.13$& 0.109  & 0.121  &0.098   & 0.173      & 0.148       & 0.149           &0.124        &0.134      & 0.111       \\[2mm]
\hline\\
\end{tabular}
\end{center}
\end{table*}

\subsection{Sub-grid gas clumping}

We also study the effect of gas clumping at very small scales, below the
resolution of our current simulations. This clumping would increase the
recombination rate. The sub-grid clumping coefficient, 
$C_{\rm sub-grid}=\langle n^2\rangle/\langle n\rangle^2$, we use is given by 
\be
C_{\rm sub-grid}(z)=27.466 e^{-0.114z+0.001328\,z^2}.
\label{clumpfact_fit}
\ee
for WMAP1 cosmology (a good fit for $8<z<40$) and
\be
C_{\rm sub-grid}(z)= 26.2917e^{-0.1822z+0.003505\,z^2}.
\label{clumpfact_fit3}
\ee
for WMAP3 cosmology (a good fit for $6<z<30$), in which case the structures form
later. These fits to the small-scale clumping factor are a more precise
version of the one we presented in \citet{2005ApJ...624..491I}. To derive it 
we used a PMFAST simulation with the same computational mesh, $3248^3$, and 
number of particles, $1624^3$, but a much smaller computational volume, 
$(3.5\,\rm h^{-1}~Mpc)^3$, and thus much higher resolution. These parameters 
correspond to particle mass of $10^3M_\odot$, minimum resolved halo mass 
of $10^5M_\odot$, and a spatial resolution of $\sim1$~kpc comoving. This box 
size was chosen so as to resolve the scales most relevant to the clumping - on 
smaller scales the gas would be Jeans smoothed, 
while on larger scales the density fluctuations are already present in our 
computational density fields and should not be included again. 

The expressions in equations~(\ref{clumpfact_fit}) and (\ref{clumpfact_fit3}) 
exclude the matter residing inside collapsed halos since these contribute to 
the recombination rate differently from the unshielded IGM. The minihalos are 
self-shielded, which results in their lower contribution to the total number
of recombinations than one would infer from a simple gas clumping argument 
\citep{2004MNRAS.348..753S,2005MNRAS...361..405I}. In principle, the
additional consumption of ionizing photons by minihalos can also be included 
as a sub-grid prescription, as we have done elsewhere \citep{MH_sim}, which 
results in a further delay of the final overlap. As discussed by
\citet{2005ApJ...624..491I}, however, the biased clustering of minihaloes
around the larger mass source halos tends to make the minihalo photon
consumption correction degenerate with the efficiency for source-halos to
release ionizing photons, so we will assume here that this efficiency
parameter takes approximate account of this effect. The larger halos, on the
other hand, are assumed to be ionizing sources, and their recombinations are
implicitly included in the photon production efficiency, $f_\gamma$, through
their escape fraction. We can neglect the gas density fluctuations associated
with suppressed sources. This gas is presumed to have been prevented by
pressure forces from collapsing out of the IGM into these suppressed halos, so
it should not contribute to the clumping factor as if it were inside the
halos. Pressure forces may also have affected the clumping of the diffuse IGM 
inside the H~II regions. Our estimate of the clumping factor, therefore,
serves to bracket the effect of IGM clumping. A fully self-consistent
treatment requires following the detailed gas dynamics coupled to the
radiative transfer, but this would require numerical resolution ($\simlt
1$~kpc resolution at $\sim100$~Mpc scale) which is currently unfeasible.    

\subsection{Simulation Cases}

\begin{table}
\caption{Simulation parameters and global reionization history results for
runs with WMAP3 cosmological parameters. Box sizes are in [$\,h^{-1}$Mpc].}
\label{summary_wmap3}
\begin{center}
\begin{tabular}{@{}lllll}\hline
                          &f250& f2000\_250S & f2000C\_250S & f250\_250S\\[2mm]
\hline
mesh                      &$203^3$& $203^3$     & $203^3$      & $203^3$\\[2mm]
box size                  &100& 35          & 35           & 35        \\[2mm]
$(f_\gamma)_{\rm large}$  &250& 250         & 250          & 250       \\[2mm]
$(f_\gamma)_{\rm small}$  & - & 2000        & 2000         & 250       \\[2mm]
$C_{\rm subgrid}$         &1  & 1           & $C(z)$       & 1         \\[2mm]
Jeans supp.               &-  & yes         & yes          & yes       \\[2mm]
\hline\\ 
$z_{50\%}$                &8.9& 10.4        & 9.7          & 9.3       \\[2mm] 
$z_{\rm overlap}$         &7.5& 7.9         & 6.9          & 7.5       \\[2mm] 
$\tau_{\rm es}$           &0.082& 0.103     & 0.096        & 0.089     \\[2mm]
\hline\\
\end{tabular}
\end{center}
\end{table}

In this paper we present a total of 15 simulations, of which 11 utilize the
WMAP1 background cosmology. Five of these have a
simulation box size of $100\,h^{-1}$~Mpc and were presented in Papers I and
II. This corresponds to a particle mass of $2.5\times10^7\,M_\odot$ and
minimum resolved halo mass of $2.5\times10^9\,M_\odot$ (requiring 100
particles or more to make sure halos are properly identified). The other 
six simulations have a smaller box size of $35\,h^{-1}$~Mpc. 
This allows for better mass and spatial resolution, corresponding to a 
particle mass of $10^6\,M_\odot$ and minimum resolved halo mass of 
$10^8\,M_\odot$. This last mass roughly corresponds to the (generally 
redshift-dependent) minimum halo mass required for its gas to be able to 
cool efficiently by hydrogen line cooling (a gas virial temperature of 
$\sim10^4$~K) \citep[e.g.][]{2001MNRAS.325..468I}. These simulations and 
their basic parameters and features are summarized in 
Table~\ref{summary}\footnote{The integrated electron-scattering optical 
depths for our large-box simulations were 
estimated incorrectly in \citet{2006MNRAS.369.1625I} and 
\citet{21cmreionpaper}, the correct, slightly lower, values are listed in 
Table~\ref{summary}}.
Note that two of our simulations, f2000\_406 and f2000\_250S\_406 have the 
higher grid resolution of $406^3$ for the radiative transfer calculation 
but are otherwise identical to simulations f2000 and f2000\_250S, 
respectively, and thus serve to study possible resolution effects. 

Additionally, we investigate the effects of varying the background 
cosmological parameters by doing four additional simulations for which we
adopt the WMAP3 background cosmology but otherwise make the exact same
assumptions about the reionization sources as the corresponding WMAP1
simulations. These simulations also impose periodic boundary conditions on
the ray-tracing, rather than the transmissive ones adopted for the simulations 
in Table~\ref{summary}. The periodic boundary conditions in the radiative 
transfer simulations are implemented by (logically) positioning each 
ionizing source at the center of the grid using the periodicity of the 
density field, before calculating its contribution to the global 
ionization rates. This results in each source having a region of influence
of the same size as our box. Since our boxes are large, they are 
optically-thick along most lines-of-sight throughout most of the evolution 
and thus the majority of the radiation is absorbed within the box. Any 
radiation that still leaves the computational volume is collected and put back 
in as a diffuse background, which boosts the effective local photoionization 
rates. The simulations using WMAP3 background cosmology are summarized in 
Table~\ref{summary_wmap3}.

The radiative transfer simulations presented in this work were run on a 
variety of computers at CITA and The University of Texas, from single- and 
dual-processor Opteron 64-bit workstations (at effective $3.6$~GHz), to 
quad-processor Itanium-2 server (at $1.3$~GHz), to a 32-processor DEC-Alpha 
machine (at $733$~MHz). The run times varied from 500 up to $\sim15,000$ 
Opteron-equivalent processor-hours for the simulations with $203^3$ mesh, 
and about 8 times longer than that for the simulations with $406^3$ mesh.  

The maximum number of sources which we ray-trace per time-step for the 
large-box [$(100\,\rm h^{-1}~Mpc)^3$ volume] simulations is up to 
$\sim3\times10^5$ for WMAP1 cosmological parameters, and up to $\sim10^5$ for 
WMAP3 cosmological parameters. The corresponding numbers for the smaller-box, 
[$(35\,\rm h^{-1}~Mpc)^3$ volume] simulations are somewhat lower, at 
$\sim1\times10^5$ for WMAP1 and $\sim1\times10^4$ for WMAP3. The net cumulative
source-halo episodes explicitly ray-traced over the course of each simulation 
is typically a few million for the large-box simulations, and a few hundred 
thousand for the smaller-box simulations.

\section{A Simple analytical toy model for self-regulated reionization}
\label{toy_model_sect}

The effect of the suppression of small-mass sources when their halos form
inside an H~II region can be illustrated by a simple, analytical toy
model. This will anticipate the phenomenon of self-regulation of reionization
and the effects of varying the photon release efficiencies of large and small
halos and the IGM clumping factor on the evolution of the mean ionized
fraction of the universe. The universe is assumed to be comprised of H~II
regions which fill a fraction $x_v$ of the total volume and H~I regions in the
remaining volume. The global average rate of change of the ionized volume
fraction with time, $dx_v/dt$, is determined by the rate of emission of
ionizing photons and the rate of recombinations (all per atom in the
universe). The volume-averaged rates of emission of ionizing photons per
collapsed atom can be different for low-mass and high-mass source haloes,
respectively, if the photon release efficiencies, $f_{\gamma,1}$ and
$f_{\gamma,2}$ per atom are different. 
In addition, the low-mass source haloes are subject to Jeans-mass filtering,
which means that the only atoms which should be counted as collapsed onto the
low-mass haloes are those associated with haloes in the {\it neutral} regions, 
which occupy a volume fraction of the universe equal to $1-x_v$. In the
simplest approximation, in which the low-mass haloes are assumed to be
distributed uniformly or randomly in space, the volume-averaged ionization
rate contributed by the low-mass haloes is $(1-x_v)f_1(t)$, if $f_1(t)$ is
the emission rate per atom if we neglect suppression. Recombinations occur
only in the ionized regions, so the volume-averaged recombination rate per
atom is the one for fully-ionized gas, $f_3(t)$, multiplied by $x_v$, where
$f_3(t)\equiv t_{\rm rec}^{-1}=C(z)n_H\alpha_B$. This yields a simple
differential equation for the mean reionization history of the IGM, 
\be 
\frac{dx_v}{dt}=f_1(t)(1-x_v)+f_2(t)-f_3(t)x_v,
\label{model_equ}
\ee
where $f_2(t)$ is the emission rate of the high-mass haloes per atom in the
universe. Equation~(\ref{model_equ}) is conveniently rewritten in a
non-dimensional form as 
\be
\frac{dx}{dy}=-x+S,
\label{model_equ_nond}
\ee
where 
\be
S\equiv\frac{f_1(t)+f_2(t)}{f_1(t)+f_3(t)},
\label{sourcefunc_def}
\ee
and 
\be
dy=[f_1(t)+f_3(t)]dt. 
\label{y_def}
\ee
The formal solution of equation~(\ref{model_equ_nond}) is given by
\be
x_v(y)=x_v(0)e^{-y}+\int_0^ydy'e^{-(y-y')}S(y'),
\label{formal_soln}
\ee
where 
\be
y=\int_0^t[f_1(t')+f_3(t')]dt'.
\ee
This solution is familiar since equation~(\ref{model_equ_nond}) is identical 
to the static equation of radiative transfer in which $x_v$ is intensity, $y$
is optical depth and $S$ is the source function. The contribution of the
$f_1(t)$ term to the integral in equation~(\ref{formal_soln}) is the
cumulative number of ionizing photons released by all the low-mass sources
over time if suppression by Jeans-mass filtering is neglected,
$\xi_{0,1}\equiv\int_0^tf_1(t')dt'$, while the integral over $f_3(t)$ is just
the cumulative total number of recombinations per atom in a fully-ionized
universe, a large number. As long as we start the time integration at some
finite cosmic time after the Big Bang (e.g. the recombination epoch), this
recombination integral remains finite. In that case, $y\gg1$ and $x_v(0)=0$,
and we can replace $x_v(t)$ by $S(t)$, yielding
\be
x_v(t)=\frac{f_1(t)+f_2(t)}{f_1(t)+f_3(t)}.
\label{hightau_soln}
\ee 

According to equation~(\ref{hightau_soln}), reionization ends at overlap epoch
$t_{\rm ov}$ such that $x_v=1$, when $f_2(t_{\rm ov})=f_3(t_{\rm ov})$, while
equation~(\ref{model_equ}) tells us that $(dx_v/dt)_{\rm ov}=0$.

The solution {\it without} suppression is given by first setting $f_1(t)=0$ in
the equations above and then replacing $f_2(t)$ by $f_1(t)+f_2(t)$:
\be
[x_v(t)]_{\rm \mbox{no supp.}}=\frac{f_1(t)+f_2(t)}{f_3(t)}.
\label{hightau_soln_nosupp}
\ee
In this case, reionization ends (i.e. $x_v=1$) when $f_1(t)+f_2(t)=f_3(t)$,
and the derivative $dx_v/dt$ is again zero. 

A comparison of the solutions above for the cases with and without suppression
shows several things. Suppression delays the completion of reionization, since
the emission of ionizing photons by the low-mass sources adds to that of the
high-mass haloes to reionize the universe and balance the recombinations at an
earlier epoch when suppression is neglected. With suppression, the overlap
epoch is determined entirely by the balance between the emission contribution
of the high-mass haloes and recombinations, independent of the contribution
from the low-mass haloes. This illustrates just what is meant by
``self-regulation'', since the overlap epoch in the presence of suppression is
{\it independent} of the efficiency for photon release by the low-mass
haloes. Apparently, if they are more efficient at releasing photons, they are
also more efficient at suppressing the formation of other low-mass sources, so
their net effect is the same. In either case, overlap is achieved by the
high-mass sources, which are free of suppression. Prior to overlap, however,
the volume ionized fraction $x_v$ is higher at every epoch when suppression is
neglected. If we were to {\it neglect} the small-mass sources altogether, we
would find that reionization is also delayed, of course. But these results
indicate that if we add the contribution of small-mass sources but {\it
  account} for their suppression, the resulting overlap will be the same as if
we neglected the small-mass sources altogether.  

Reionization is {\it extended} by the presence of low-mass haloes, however,
relative to the case with no low-mass haloes. According to
equations~(\ref{hightau_soln}) and (\ref{hightau_soln_nosupp}), $x_v$ is
higher at all times when $f_1\neq0$ than when $f_1=0$, because
$(f_1+f_2)/(f_1+f_3)>f_2/f_3$, since $f_3>f_2$ at all times prior 
to the instant of overlap (i.e. since $x_v<1$ before overlap, when $x_v=1$). 
This means that the effect of adding low-mass haloes would be to
{\it increase} the integrated electron-scattering optical depth $\tau_{\rm
  es}$ {\it without} changing the overlap redshift.     

Similar simple analytical models, based on equation~(\ref{model_equ}), with 
some variations, has been explored in several recent works, albeit 
none derived the analytical solution presented in 
equations~(\ref{hightau_soln}) and (\ref{hightau_soln_nosupp})
\citep{2003ApJ...595....1H,2003ApJ...586..693W,2004ApJ...610....1O,
2005ApJ...634....1F}. These studies largely reached analogous conclusions 
to ours, namely that the Jeans-mass suppression extends reionization, but 
only rarely, if ever, does this lead to a non-monotonic evolution of the 
ionized fraction, or to a double reionization.

In reality, the reionization history is more complicated than our toy model
suggests. As we shall see, the source haloes and their H~II regions are
spatially-clustered and thus biased relative to the matter distribution, so
our assumptions above of uniformly distributed low-mass haloes and uniform gas
density are not correct. We will revisit this issue in \S~\ref{results_sect}
and the Appendix and show how we can improve the toy model with hindsight from
our detailed simulation results. None of the above previously-published 
semi-analytical reionization models above include the effects of bias.

\section{Simulation Results}
\label{results_sect}

\subsection{Mean reionization history milestones}

We start by examining the mean global reionization histories derived 
from our simulations. These can be characterized by several basic 
parameters. The first of these parameters is the epoch of overlap, 
$z_{\rm ov}$, which we define as the time when the mass-weighted ionized 
fraction of the gas first surpasses 99\%. This also quantifies the overall 
duration of reionization, since its start is determined by when the first 
resolved halos form in our simulations, and thus is fixed by the structure 
formation alone. The second global parameter is the total electron scattering 
optical depth, $\tau_{\rm es}$, integrated from the beginning of reionization
to the present. The third and final parameter is the redshift 
when 50\% of the gas mass is ionized for the first time. This is 
of direct interest for observations since this half-ionization point is a 
good indicator of the epoch when the fluctuations of the redshifted 21-cm 
emission from neutral hydrogen reach their maximum \citep{21cmreionpaper}.

\begin{figure*}
\includegraphics[width=3.2in]{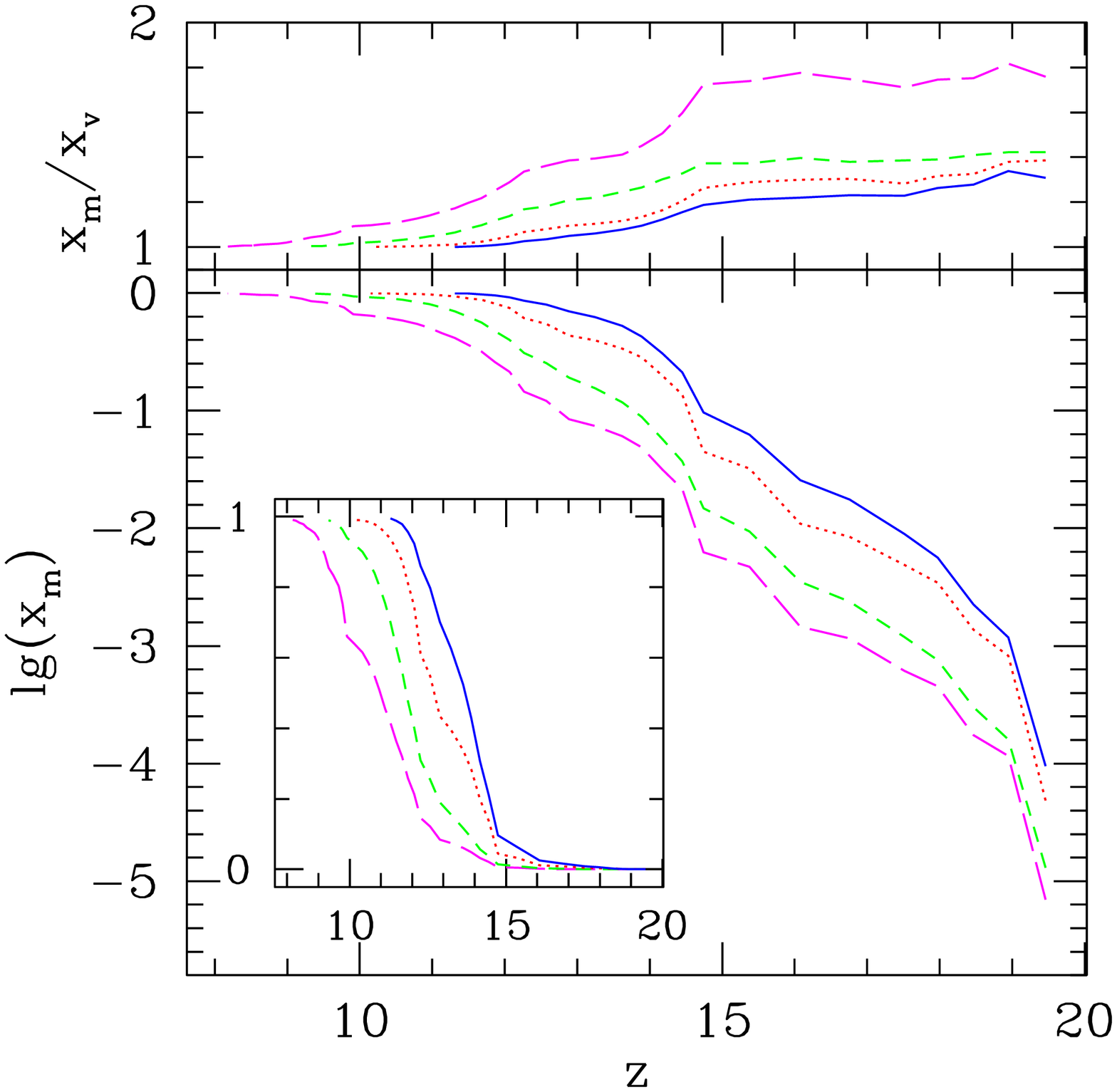}
\includegraphics[width=3.2in]{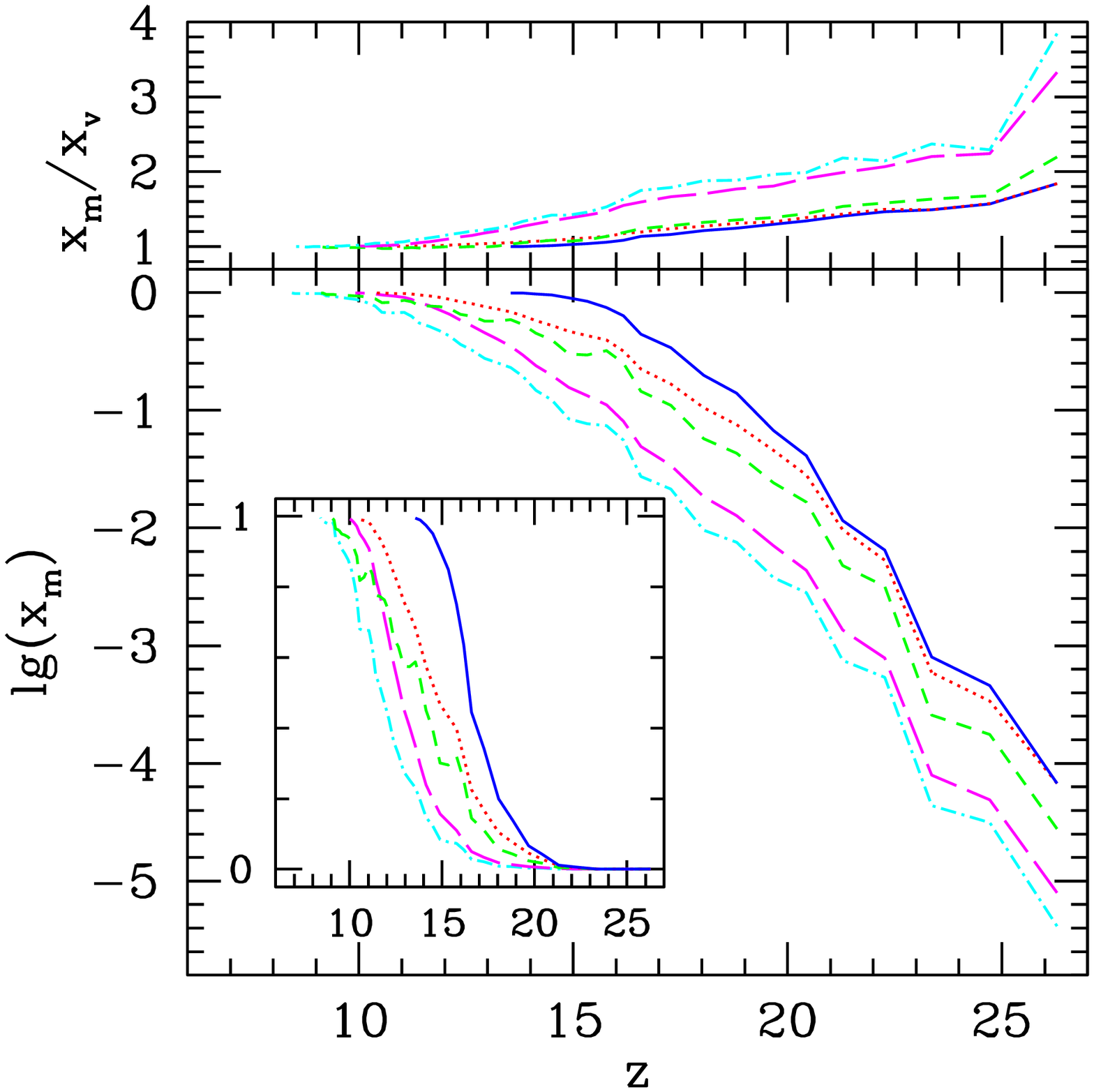}
\caption{Mean reionization histories: (a)(left) $100\,h^{-1}$~Mpc box runs: 
(bottom panel) redshift evolution of the mass-weighted ionized fraction, 
$x_m$ for f2000 (solid, blue), f2000C (dotted, red), f250 (short-dashed, 
green), and f250C (long-dashed, magenta). (top panel) Corresponding ratios 
of mass-weighted and volume-weighted ionized fractions, which are equal to 
the mean density of the ionized regions in units of the mean density of the 
universe, and (b)(right) $35\,h^{-1}$~Mpc box runs: f2000\_250 (solid, blue), 
f2000\_250S (dotted, red), f2000C\_250S (short-dashed, green), f250\_250S 
(long-dashed, magenta), and f250C\_250S (dot-short dashed, cyan). Insets: 
the same reionization histories, but in linear scale.
\label{fracs_35Mpc}}
\end{figure*}
For all the large-box simulations of comoving size $100\,h^{-1}$~Mpc on
a side the first resolved halo forms, and thus reionization starts, at $z\sim20$ 
(16) in WMAP1 (WMAP3) cosmology. For the smaller-box simulations of size 
$35\,h^{-1}$~Mpc this occurs earlier, at $z\sim30$ for WMAP1 cosmology and at 
$z\sim22$ for WMAP3 cosmology. This earlier start for smaller simulation volume 
is due to the hierarchical nature of CDM structure formation, whereby the smaller 
halos form earlier. Our N-body simulations have the same number of particles, thus 
the smaller boxes have correspondingly higher mass resolution, and hence the 
first halos form earlier. Among our simulations, the earliest epoch of overlap 
is $z_{\rm ov}=13.5$, achieved in the (somewhat artificial) case f2000\_250, 
i.e. when small-mass sources are never suppressed by Jeans mass filtering,
while at the same time they are highly efficient at producing ionizing
photons. The  
combination of these two properties results in a very large photon emissivity, 
and as a consequence, a very fast reionization and too early overlap. In the 
more realistic cases, in which the small-mass source suppression is included, 
the overlap occurs significantly later, between redshift $z=10.4$ (f2000\_250S) 
and 8.4 (f250C\_250S). For the large-volume simulations, the redshifts of 
overlap are similar, ranging from $z=11.3$ (f2000) to $z=8.2$ (f250C). 
The reason for these similar end-of-reionization times despite the lack of 
small-mass sources in the large-volume simulations is that most of these small 
sources become suppressed by the time of overlap, and thus, in either case, the 
reionization is brought to an end by the same large-mass sources. However, at 
early times there are significant differences between the results for the two 
box sizes. As we noted above, the small-mass sources start forming significantly 
earlier, thus models which resolve these sources have higher ionized fraction 
at early times. This results in higher integrated optical depths when small-mass 
sources are resolved. For the WMAP1 cases, these range from $\tau_{\rm es}=0.173$ 
(f2000\_250) when the small-mass sources are high-efficiency but not suppressed, 
to $\tau_{\rm es}=0.148$ (f2000\_250S) when suppression is taken into account, 
to $\tau_{\rm es}=0.111$  (f250C\_250S) when low-mass sources have the same 
efficiency as the high-mass ones and sub-grid IGM clumping is taken into account, 
all of which are roughly within 1-$\sigma$ from the WMAP1 estimate, 
$\tau_{\rm es}=0.17\pm0.04$. In the $100\,h^{-1}$~Mpc box simulations, the 
small-mass sources are not present and thus reionization starts later. This 
results in somewhat lower values of the integrated optical depth, ranging from 
$\tau_{\rm es}=0.130$ (f2000), which is nevertheless still in agreement with 
the WMAP1 measurement, to $\tau_{\rm es}=0.098$ (f250C), which is a bit low. 

Increasing the resolution to $406^3$ grid has only a modest effect on the 
reionization histories. In these cases the underlying density field is resolved 
better, which results in an increased recombination rate, and a slightly more 
extended reionization history.  

\begin{figure}
\includegraphics[width=3.2in]{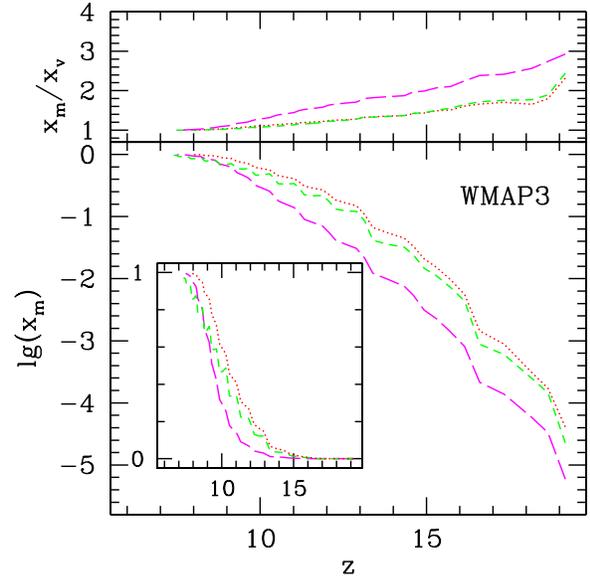}
\caption{Mean reionization histories: $35\,h^{-1}$~Mpc box runs with 
WMAP3 background cosmology: f2000\_250S (dotted, red), f2000C\_250S 
(short-dashed, green), and f250\_250S (long-dashed, magenta).
Inset: the same reionization histories, but in linear scale.
\label{fracs_35Mpc_wmap3}}
\end{figure}

\subsection{WMAP1 vs. WMAP3 and ionizing source efficiencies}
\label{wmap13_sect}

Within the WMAP3 background cosmology, structure formation occurs later 
and all the evolution shifts to correspondingly lower redshifts. For the
same efficiencies, overlap is now reached at redshifts between 
$z_{\rm ov}=7.9$ (f2000\_250S) and $z_{\rm ov}\sim7$ (f2000C\_250S), in 
rough agreement with the data from high-redshift galaxies and SDSS QSO's, 
which indicate a tail-end of reionization 
at $z_{\rm ov}\sim6-7$ \citep[e.g.][]{2002AJ....123.1247F,2003AJ.126..1W,
2004ApJ...617L...5M}. The resulting integrated electron scattering optical 
depths are correspondingly lower, at $\tau_{\rm es}=0.082-0.103$, in complete
agreement  
with the WMAP3 constraints. These results confirm that the delay in structure 
formation due to the lower values of $\sigma_8$ and $\Omega_0$, higher value 
of $h$ and the slight tilt of the primordial power spectrum found by WMAP3 
results yield a decrease of the integrated optical depth from the WMAP1 value 
of $\sim0.17$ to the WMAP3 value of $\sim0.09$, as was previously predicted 
analytically \citep{2006ApJ...644L.101A}. This analytical estimate predicted 
that the change of the 
cosmological parameters from the WMAP1 to WMAP3 values would result in a
delay of reionization by a factor of 1.4 in $(1+z)$ and a corresponding lower 
optical depth by a factor of $1.4^{3/2}\sim1.7$. These predictions are nicely 
confirmed by our current simulations, which yield a decrease in the redshift 
of 50\% ionization by a factor of $\sim$1.3-1.4, and in the overlap redshift by 
a factor of $\sim$1.3, while the $\tau_{\rm es}$ values decrease by factors of 
$\sim1.4\approx1.3^{3/2}$. The correction factors are slightly smaller than 
the ones derived analytically. These differences are due mostly to the use of 
periodic boundary 
conditions in our WMAP3 simulations (which ensures that no photons are lost 
through the simulation box boundary), vs. our use of transmissive boundary 
conditions in the earlier simulations. This yields a slightly faster evolution 
at the late stages of reionization, and correspondingly earlier overlap. 

There has been a recent claim \citep{2006astro.ph..5358P} that the delay in 
structure formation due to the different background cosmological parameters 
derived by WMAP3 is not in fact sufficient to account for the reduction of the 
derived optical depth, once the feedback and radiative transfer effects are 
accounted for. This work claimed, instead, that the WMAP3 cosmology requires a 
more top-heavy IMF in order to match the new value of the integrated optical 
depth. Our simulations show conclusively that this is not the case.  

Our results show that, while a very top-heavy IMF is not required, ionizing 
sources nonetheless must have been fairly efficient photon producers. 
This means that, compared to the present-day observed galaxies, the 
high-redshift galaxies must have had either a moderately top-heavy IMF, a 
higher star formation efficiency, a higher photon escape fraction, or some 
combination of these. 

Recently, \citet{2006astro.ph..4177Z} performed a simulation similar to the 
ones we presented in Papers I and II. This work used an approach similar to
ours in coupling a radiative transfer scheme to the results of an N-body
simulation of the density field. They adopted WMAP1 cosmology parameters, 
but utilized a smaller simulation volume and a different radiative transfer 
method. The sources resolved in these simulations have minimum mass 
$2\times10^9\,M_\odot$, very similar to the resolution of our large-box 
simulations. In terms of ionizing source efficiencies, they assumed that all 
the sources only produce a cumulative one photon per every atom in the
universe by redshift $z\sim6.5$. This corresponds to a much lower source
emissivity than the ones we have assumed. Such a model predicts quite late 
reionization and a very low integrated 
electron scattering optical depth of $\tau_{\rm es}=0.06$. The final overlap 
is achieved by $z\sim6$ only if the number of recombinations per atom 
integrated over time is negligible. As we noted above, adopting the WMAP3 
cosmology for these simulations would inevitably push the redshift of overlap 
down to $\sim4$ and the optical depth down to 0.035 (and even further down if 
recombinations are included), in clear disagreement with the observations. 
This again demonstrates that relatively high photon production efficiencies 
are required for the high redshift sources, similar to the ones we adopt here.  

\begin{figure}
\vspace{-0.5in}
\begin{center}
\includegraphics[width=\textwidth]{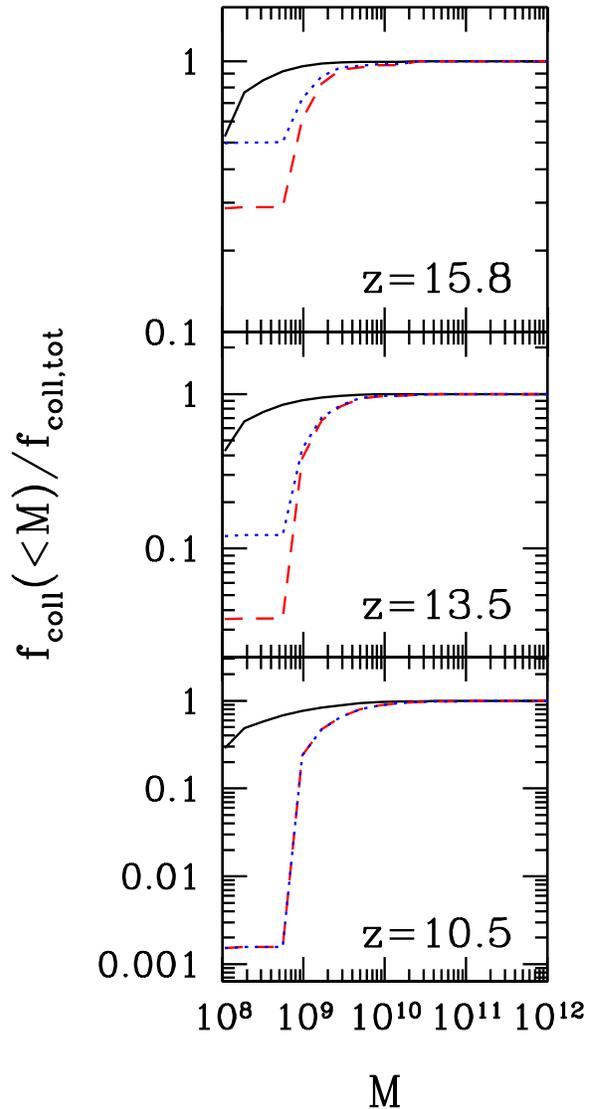}
\vspace{-0.7in} \caption{
Effect of Jeans-mass filtering on the collapsed mass fraction of halos in our
simulations. Shown are the cumulative collapsed mass fractions in halos of
mass less than $M$ (in units of the respective total source halo collapsed
fraction at that time) vs. $M$ for all source halos (solid, black), and the 
unsuppressed source halos for cases f2000\_250S (short-dashed, red) and 
f250\_250S (dotted, blue) at three redshifts, as labelled.  
\label{fcoll_supp}}
\end{center}
\end{figure}

\begin{figure}
\begin{center}
\includegraphics[width=3.5in]{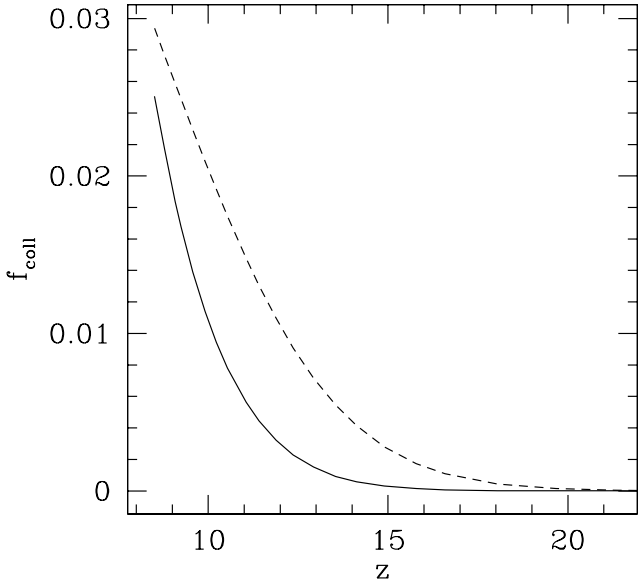}
 \caption{Collapsed fractions in low-mass halos 
($10^8\leq M/M_\odot\leq 10^9$; dotted) and high-mass halos ($M/M_\odot>10^9$; 
solid) for the simulation with $(35\,\rm h^{-1}Mpc)^3$ volume and WMAP1
   cosmology.  
\label{fcoll_tot}}
\end{center}
\end{figure}

\subsection{Globally-averaged reionization histories}

The full reionization histories (mass-weighted ionized fractions vs. redshift)
from all of our simulations are shown in Figs.~\ref{fracs_35Mpc} and 
\ref{fracs_35Mpc_wmap3}. The character of reionization is similar in all 
cases, regardless of the simulation volume or the background cosmology. In 
all cases reionization is clearly inside-out, with the high-density regions 
being ionized on average earlier than the low-density regions, as shown by
the ratio of the mass-weighted to the volume-weighted ionized fraction,
$x_m/x_v$, which reflects the mean density of the ionized bubbles compared to
the mean density of the universe, as reported previously in 
\citep{2006MNRAS.369.1625I}. These ratios are
plotted in the top panels of Figs.~\ref{fracs_35Mpc} and \ref{fracs_35Mpc_wmap3}. 
In all cases and at all times the ionized regions are overdense, more 
significantly so at early times. This inside-out character of the reionization 
process is more pronounced for the small-box simulations which have higher 
spatial resolution and thus follow the underlying density field more closely.

The evolution is always monotonic, and in no case we do see a double 
reionization or even a modest temporary decrease of the ionized fraction. 
The evolving source efficiencies (from a high one early-on to a low one 
later), small-scale gas clumping (from low values early to higher values 
later), and the Jeans-mass filtering of sources all only slow the 
process  down and extend it in time, but never manage to reverse it even 
temporarily. The high-resolution, small-box simulations find somewhat 
more extended reionization than the large-box simulations, due to the 
earlier onset of reionization in these cases, combined with the 
suppression of small-mass sources. 

\subsection{Self-regulation}

\begin{figure*}
\begin{center}
\includegraphics[width=3.2in]{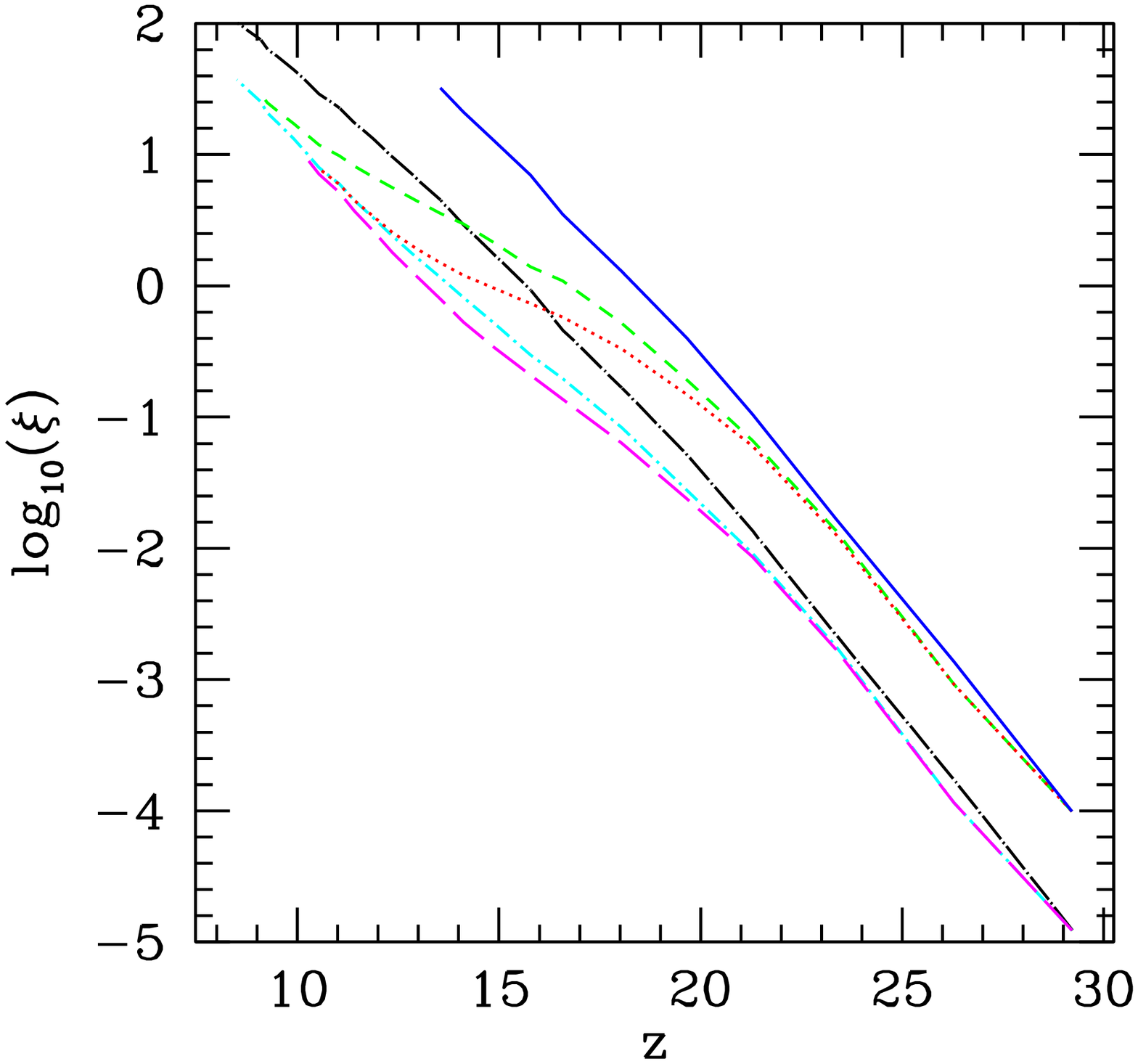}
\includegraphics[width=3.2in]{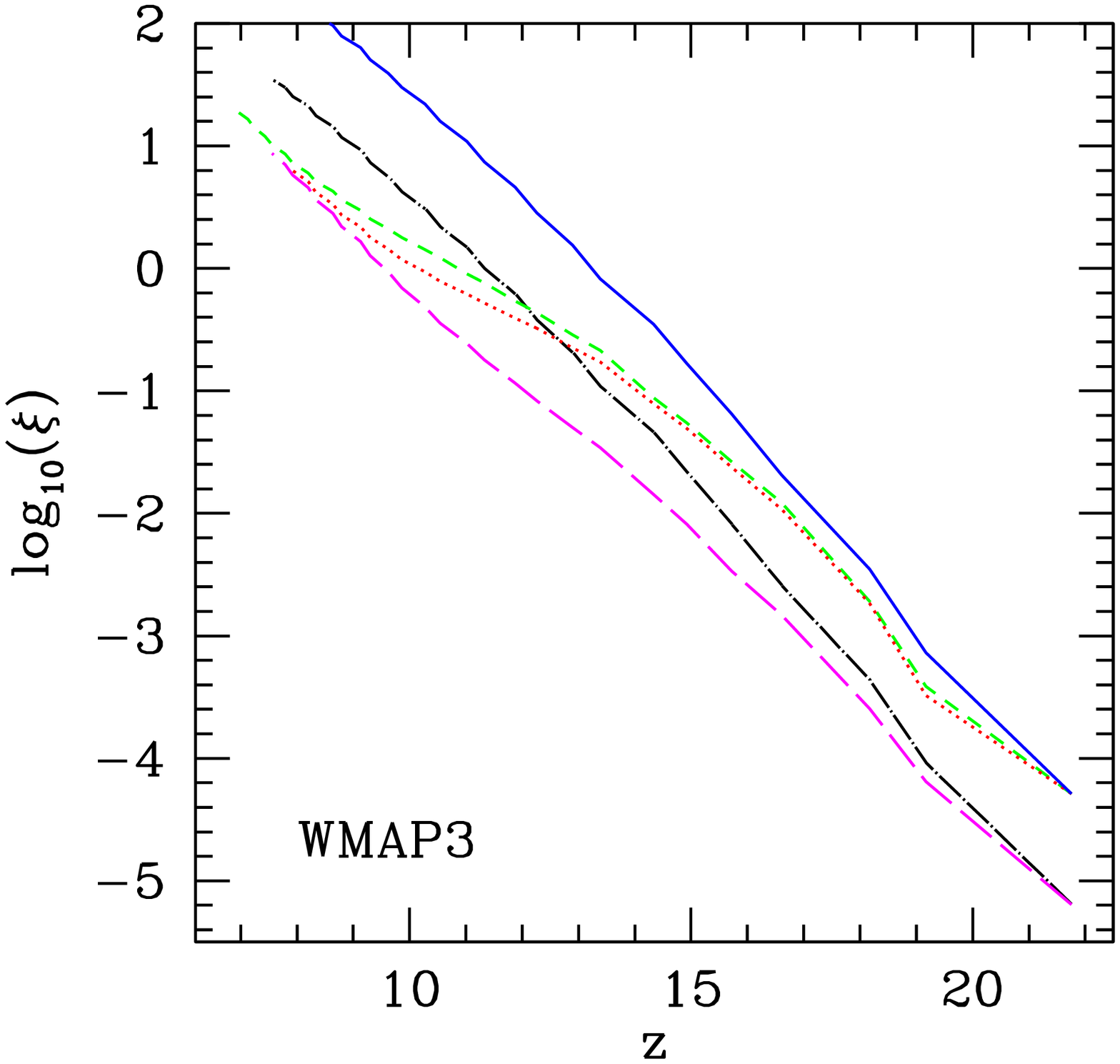}
\caption{The cumulative number of ionizing photons per total atom, $\xi$, 
emitted by all sources in the $35\,h^{-1}$~Mpc box simulations with WMAP1
(left) and WMAP3 cosmology (right). Same notation as in
Figure~\ref{fracs_35Mpc} (b) and Figure~\ref{fracs_35Mpc_wmap3}, 
respectively. For reference we also show the total $\xi$ for  
f250\_250S and f250C\_250S simulations if no sources were suppressed 
(dot-long dashed, black). 
\label{photons_35Mpc}}
\end{center}
\end{figure*} 

One interesting and important consequence of the low-mass source suppression
is that the time of overlap becomes fairly insensitive to the efficiency of
the small-mass sources, since these are almost completely suppressed by the
reionization end. Furthermore, in cases when the small-mass sources are
more efficient photon emitters, their H~II regions expand quickly, thereby
suppressing further source formation in a larger volume than less efficient
sources do. This is illustrated in Figure~\ref{fcoll_supp}, where we show the
cumulative source halo collapsed fraction with and without suppression,
normalized to the corresponding totals, at
three different redshifts - early, middle and late in the evolution. During
the early evolution, the high-efficiency of the small sources helps suppress
a significantly larger fraction of them, by a factor of a few or more, compared to
the low-efficiency case. At late times, the level of suppression becomes very
large and essentially independent of the small-source efficiency. By then most
of the volume is ionized, and due to the strong bias in their spatial 
clustering, concentrating them in the ionized regions, the effect on the 
small-mass sources is even more dramatic than simply the ionized volume would
suggest. This suppression notably slows down the further expansion of the
ionized bubbles, thus resulting in a very efficient self-regulation of
reionization and a relative insensitivity of the global evolution to the
assumed small-mass source efficiency. For example, if we compare the cases
f2000\_250S and f250\_250S (with WMAP1 parameters), we find that the ratio of
their mass ionized fractions starts at $\sim7-8$ early-on ($z>20$), decreasing
to $\sim3$ at $z\sim15$ and to $\sim1.5$ or less for $z<13$. The assumed
small-mass source efficiency is 8 times larger in the former than in the
latter case, while the larger, unsuppressible sources have the same efficiency
in the two cases. This shows that during the very early evolution the ratio of
the ionized fractions matches the ratio of the number of photons produced, but
then the ionized fractions ratio quickly decreases as the evolution progresses
and the low-mass source suppression and the corresponding self-regulation
start to dominate the reionization process. With the WMAP3 background
cosmology we observe the same generic behaviour, except that it is shifted to
later times, as we discussed above.      

\begin{figure*}
\includegraphics[width=3.2in]{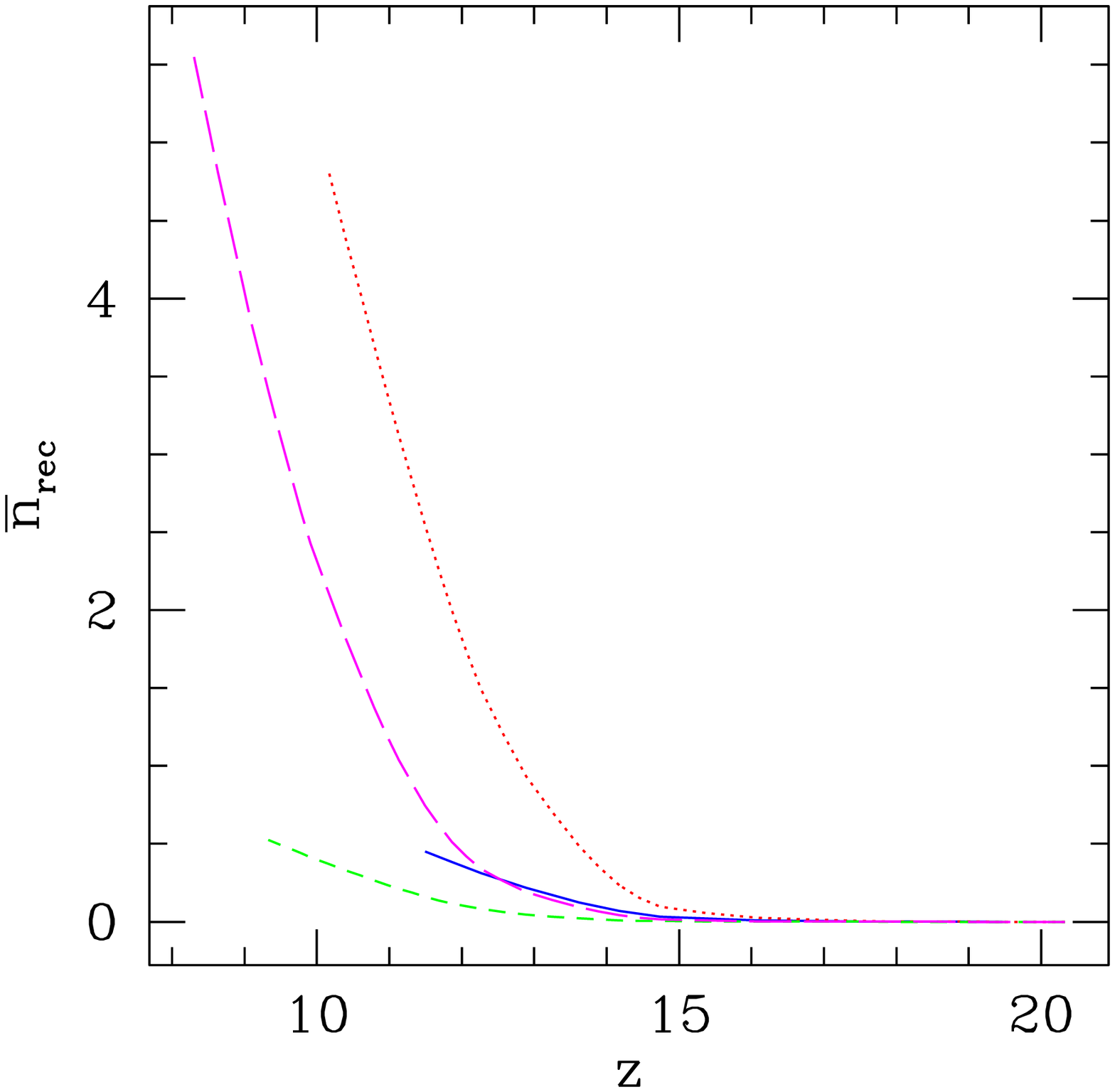}
\includegraphics[width=3.2in]{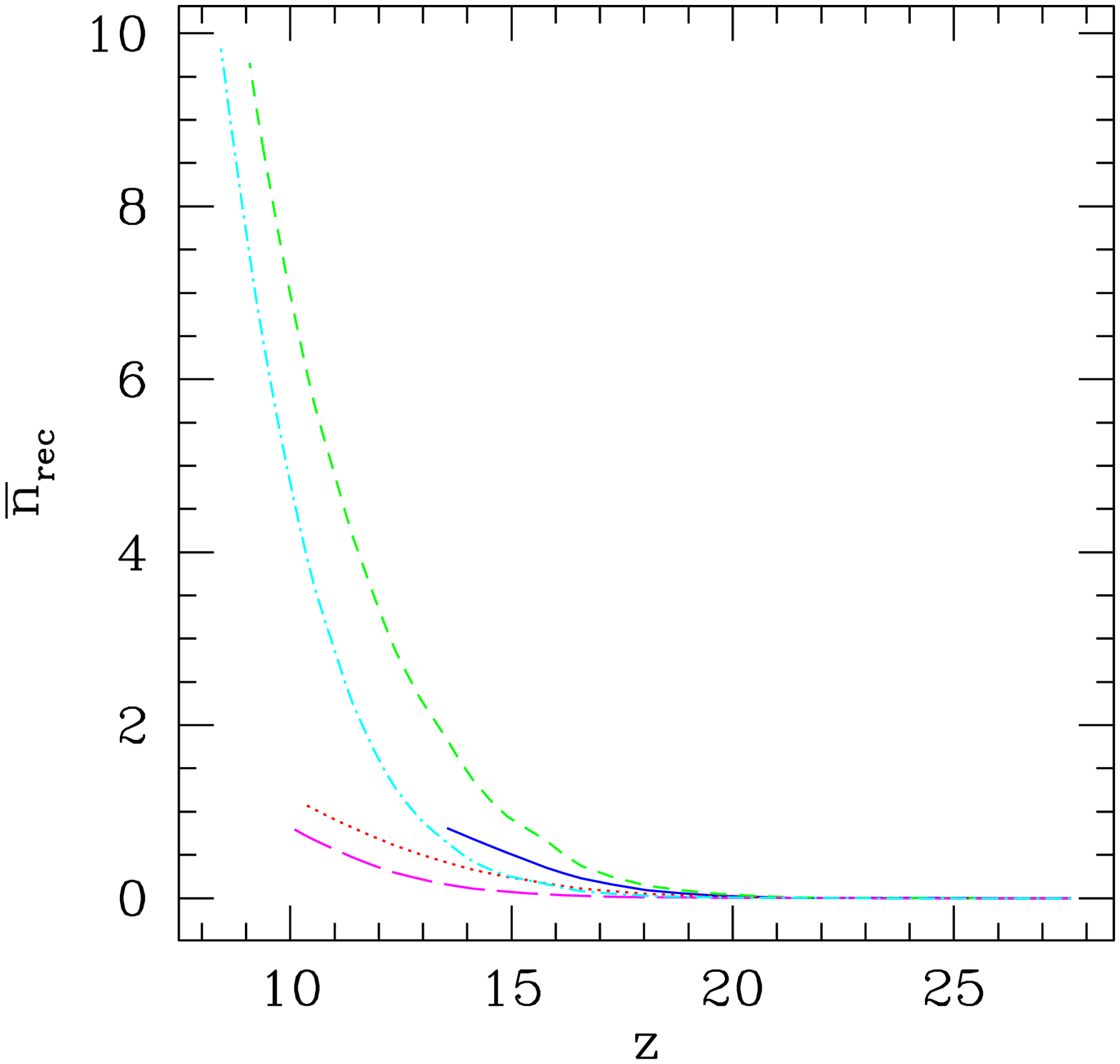}
\caption{Cumulative number of recombinations vs. redshift for: (a)(left) 
$100\,h^{-1}$~Mpc box runs, f2000 
(solid, blue), f2000C (dotted, red), f250 (short-dashed, green), and f250C 
(long-dashed, magenta), and (b)(right) $35\,h^{-1}$~Mpc box runs,
f2000\_250 (solid, blue), f2000\_250S (dotted, red), f2000C\_250S 
(short-dashed, green), f250\_250S (long-dashed, magenta), and f250C\_250S 
(dot-short dashed, cyan).
\label{ave_rec_35Mpc}}
\end{figure*}

In Figure~\ref{fcoll_tot} we show the total collapsed fractions in low-mass
and high-mass halos for the $35\,\rm h^{-1}Mpc$ computational box with WMAP1
parameters. At early times the collapsed fraction in small-mass halos rises
faster and without suppression they dominate the large-mass halos at all
times. The collapsed fraction in high-mass halos rises exponentially below
$z=15$. The collapsed fractions in the two mass ranges become comparable at
$z\sim8$. When suppression is included, large-mass halos begin to dominate the
reionization progress much earlier, after $z\sim14-15$, depending on the 
small-mass source efficiency, as shown in Figure~\ref{fcoll_supp} above.
 
In Figure~\ref{photons_35Mpc} we show the evolution of the cumulative number
of ionizing photons, $\xi$, emitted by all sources in our $35\,h^{-1}$~Mpc 
simulation volume, for both the WMAP1 and the WMAP3 cases. The suppression of 
sources due to radiative feedback clearly has a quite dramatic effect on the 
cumulative $\xi$, and can reduce it by up to 2 orders of magnitude in some
cases compared to the case when no sources get suppressed. The effect is
stronger when the small-mass sources are more efficient at producing ionizing
photons, again demonstrating the self-regulated nature of reionization - more
efficient sources suppress more, partially compensating for the higher
efficiency. The emissivity per unit time and per baryon  in the universe is 
given by $d\xi/dt$. This derivative is strictly positive at all times since 
$d\xi/dt=(d\xi/dz)(dz/dt)$, $dz/dt<0$ and based on Fig.~\ref{photons_35Mpc} 
also $d\xi/dz<0$), so the emissivity per baryon rises with time in all models 
we consider here .

\begin{figure}
\includegraphics[width=3.2in]{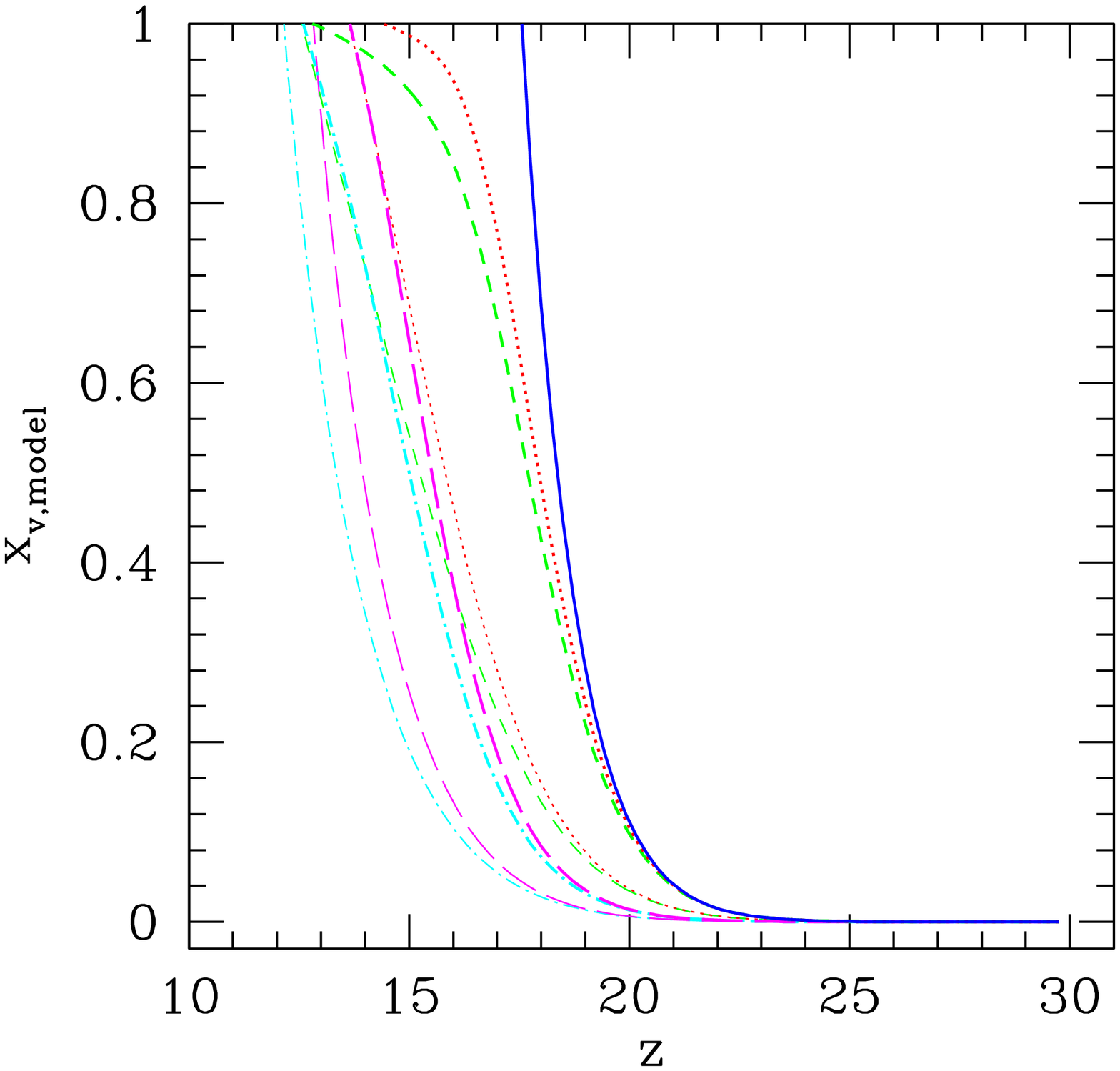}
\caption{Reionization histories based on a simple analytical model which
incorporates sources with different efficiencies, source suppression, and 
evolving gas clumping. Notation is the same as in Fig.~\ref{fracs_35Mpc} 
(right panel). Thin lines correspond to the cases with empirical ``bias'' 
included (see Appendix for details). 
\label{xv_model_fig}}
\end{figure}

In Figure~\ref{ave_rec_35Mpc} we show the cumulative number of recombinations
per atom in our computational volume. The number of recombinations is initially 
small since a very small fraction of the volume is ionized. This is despite the 
fact that these first H~II regions are highly-overdense, and thus their 
recombination times are short. At later times, the recombinations become more 
important, consuming 0.5-1 additional photon per atom when sub-grid gas 
clumping is ignored (which underestimates the recombinations) and up to 5-10
additional photons per atom when sub-grid clumping is included. This demonstrates 
that recombinations play an important role during reionization and should not be 
ignored in any simulations or analytical models.

\begin{figure*}
\includegraphics[width=3.2in]{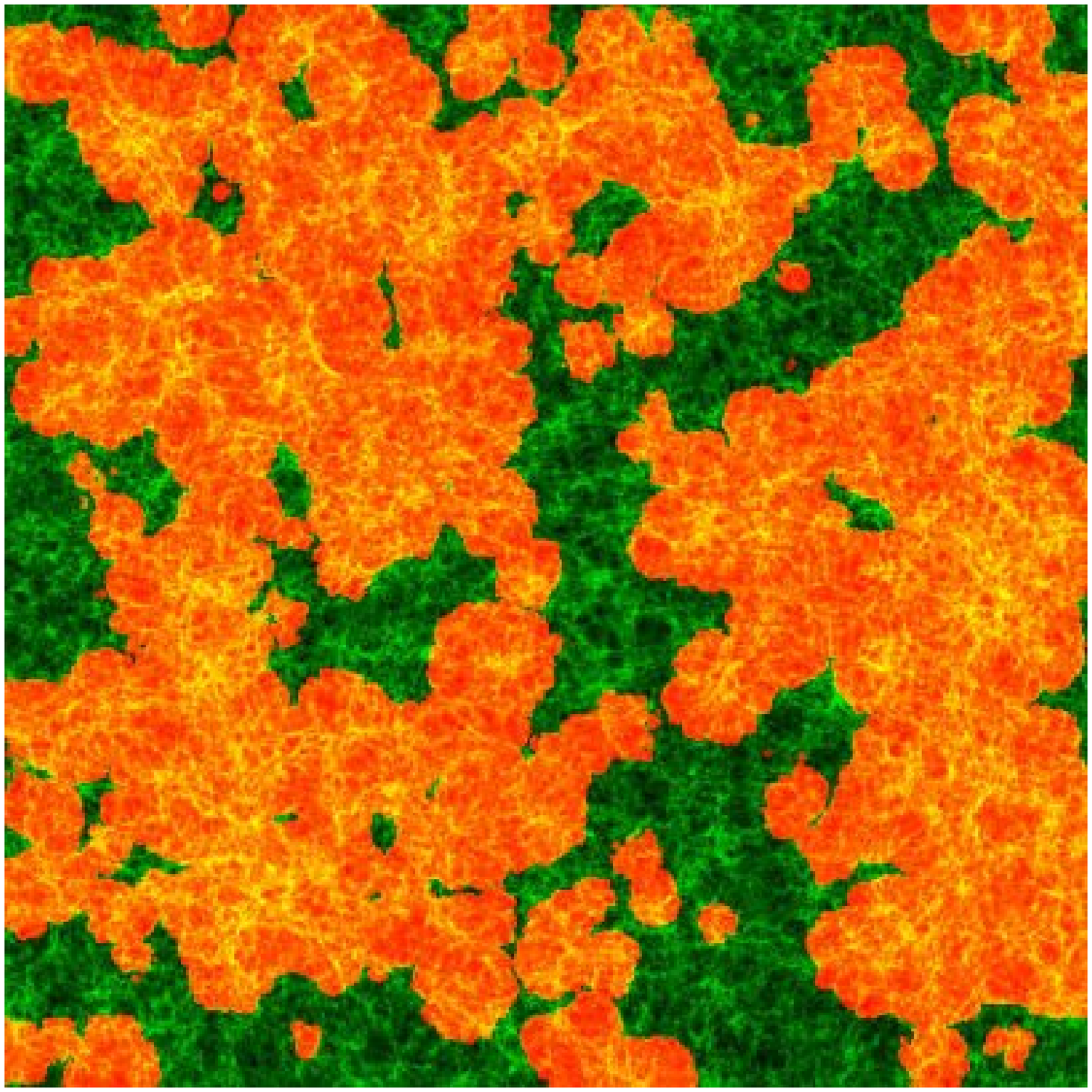}
\includegraphics[width=3.2in]{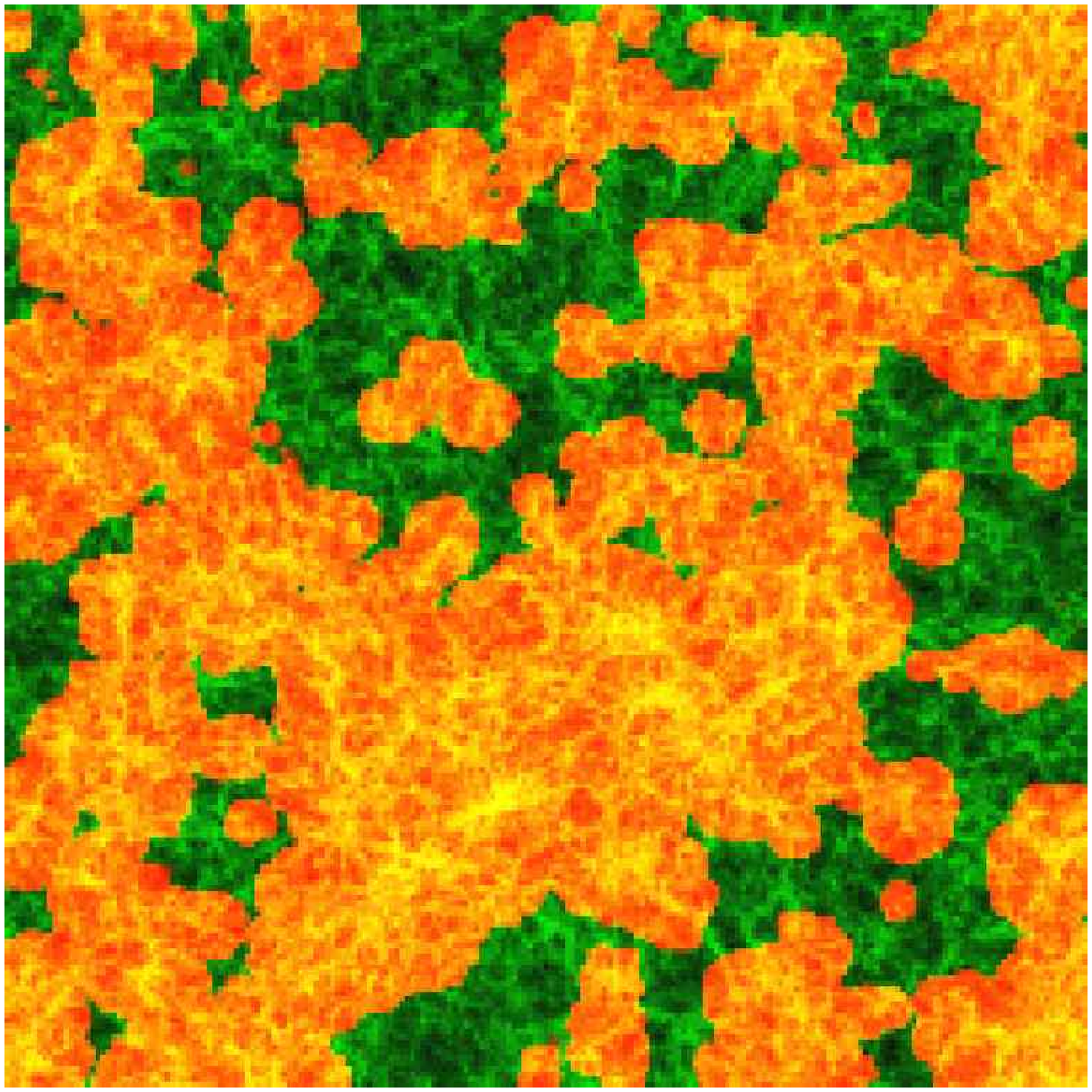}
\caption{Spatial slices of the ionized and neutral gas density from
  simulations (a)(left) f2000\_406 ($100\,h^{-1}$~Mpc
  comoving) at redshift $z=12.9$, and (b)(right) from f2000\_250 
  ($35\,h^{-1}$~Mpc comoving) at $z=16.2$. Both are at 
similar ionization fraction by mass of $x_m\sim60\%$. Shown are the density 
field (green) overlayed with the ionized fraction (red/orange). The cosmic 
web of structures is visible in both the neutral and ionized regions.   
\label{images}}
%\end{figure*}

%\begin{figure*}
\includegraphics[width=3.2in]{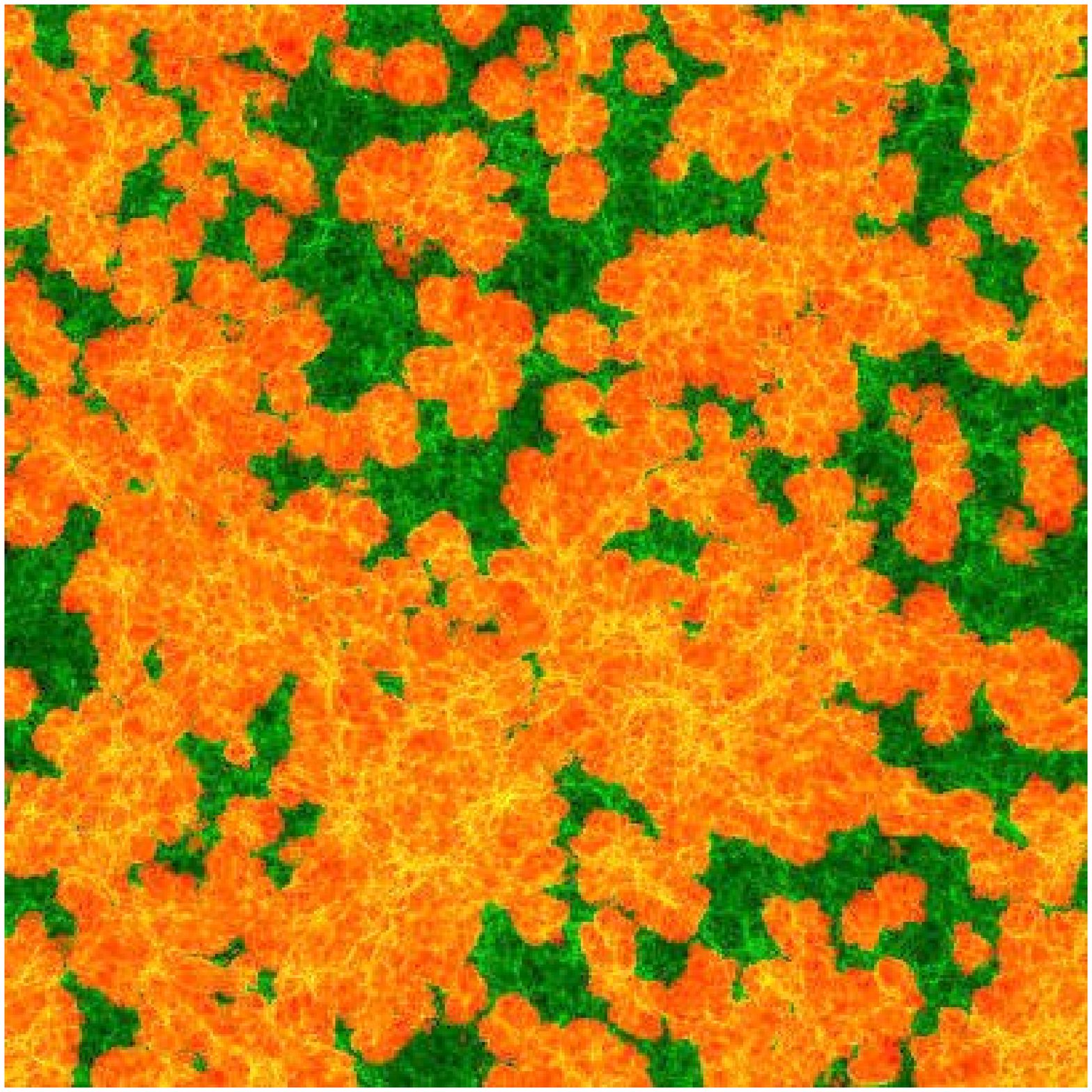}
\includegraphics[width=3.2in]{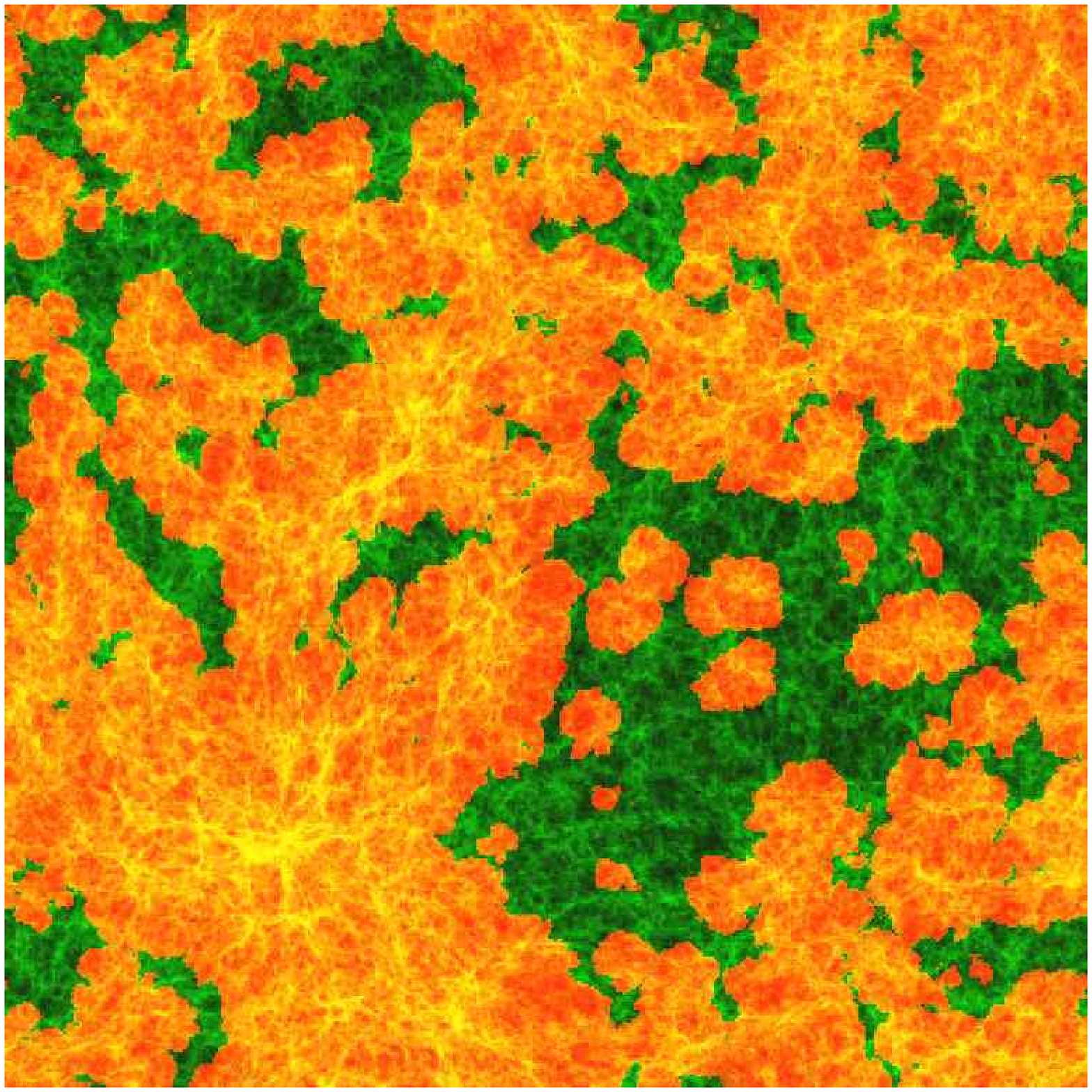}
\caption{Spatial slices of the ionized and neutral gas density from
  simulations (a)(left) f2000\_250S\_406 ($35\,h^{-1}$~Mpc
  comoving) at 
redshift $z=14.1$, and (b)(right) from f2000\_250S\_406 ($35\,h^{-1}$~Mpc
comoving) with WMAP3 
parameters at $z=10.1$. As in Figure~\ref{images}, these images 
are also at ionization fraction by mass of $x_m\sim60\%$. Shown are 
the density field (green) overlayed with the ionized fraction 
(red/orange).
\label{images2}}
\end{figure*}

Some of these trends for the basic self-regulation of reionization were
roughly anticipated by our analytical toy model in
\S~\ref{toy_model_sect}. That model neglects several effects which our
detailed simulations show are quite important, however. These include the
inhomogeneous gas density inside the ionized regions and the clustered
formation of source haloes in time and space, which biases their distribution
and that of the H~II regions relative to the underlying matter density
field. In order to account for some of these neglected effects in an
approximate way, to gain further insight and use the toy model quantitatively,
we have incorporated the results of our simulations to improve the   
simple analytical model described in \S~\ref{toy_model_sect} (see Appendix for
details). In this improved model the volume-weighted ionized fraction $x_v$
evolves according to: 
\be 
\frac{dx_v}{dt}=b_1(t)f_1(t)(1-x_v)+f_2(t)-f_3(t)x_v,
\label{model_equ0}
\ee
where $b_1(t)=(1-x_v^n)/(1-x_v)$ is the correction factor which accounts for
the fact that low-mass halos are concentrated toward the ionized regions
rather than being distributed uniformly, or randomly in space. The photon
emission rate $f_1(t)=f_{\gamma,\rm small}(f_{\rm coll,small}/\Delta t)$ is
the mean photon emissivity per atom coming from the small-mass sources based
on our modelling of sources in the radiative transfer simulations, where
$\Delta t$ is the time between two time slices ($\sim 20$~Myr), over which we 
hold the underlying gas density field and halo population fixed while we 
integrate the radiative transfer and the non-equilibrium ionization balance 
rate equations. In the toy model in \S~\ref{toy_model_sect}, the source 
suppression by Jeans-mass filtering is assumed to be limited to the ionized 
volume, with source halos randomly, or uniformly, distributed in space, thus 
the unsuppressed fraction is proportional to the neutral volume fraction,
$1-x_v$ (i.e. $n=1$ in eq.~[\ref{model_equ}]). We also consider a second 
case, where we empirically model the effects of halo spatial clustering bias 
by setting $n=0.1$ in equation~(\ref{model_equ}) (see Appendix). The second term, 
$f_2(t)=f_{\gamma,\rm large}(f_{\rm coll}/\Delta t)$ is the same as $f_1(t)$ but 
for the larger, unsuppressible sources. Finally, the last term, 
$f_3(t)=1/t_{\rm rec}=C(z)n_H\alpha_{\rm H}$ is the rate of recombinations per 
atom (assuming fully-ionized gas within the H~II regions at the mean density). 
Recombinations only occur in the ionized volume, and so the volume-averaged rate 
is just proportional to
$x_v$. In reality, the H~II regions are overdense on average, which we ignore
here, and thus our model somewhat underestimates the effect of recombinations.

Solutions for the evolution of the mean ionized fraction based on our toy
models are plotted in Figure~\ref{xv_model_fig} for the small-box simulation
cases in Table~\ref{summary}. These curves show that when low-mass sources and
their suppression are included, but bias is neglected ($n=1$), the end of
reionization is nearly the same, regardless of the efficiency adopted for
photon release per low-mass halo atom (see curves for cases f2000\_250S and
f250\_250S). When bias is incorporated ($n=0.1$) there is a very modest trend
toward an earlier end of reionization when the low-mass source efficiency is
higher. When bias is neglected, the reionization evolution tends to be too
fast compared to the simulations; bias serves to slow reionization down
because a larger fraction of the low-mass sources are suppressed at any given
time than the neutral volume fraction $1-x_v$, since these halos are
concentrated in the H~II regions. The effect of gas clumping is to delay the
end of reionization, but since clumping grows with time and the contribution from
recombinations in the ionized regions is small at early times, soon after the
sources turn on, the evolution at early times is similar with and without
clumping.   

\subsection{Effect of small-mass sources on the reionization morphology}

The morphology of the H~II regions during reionization shows many common
features, but also some interesting variations (Figs.~\ref{images} and
\ref{images2}). In general, H II regions persist and grow over time, as new
sources form to add their ionizing luminosity to the total emission rate
inside each ionized volume. In all cases reionization occurs inside-out, 
i.e. the denser 
regions around the high-density peaks are ionized first. Locally, at any 
given scale the ionized bubbles percolate at very different times. This 
local overlap leads to quite large (of size 10~Mpc comoving, or larger) 
ionized regions fairly early, well before the whole universe is ionized. 
This explains why previous, small-box simulations failed to see the gradual
build-up of the ionized volume of the universe over time, prior to
the final overlap epoch, and mistakenly found that reionization is a very
rapid transition.  At all times prior to overlap, there are also many
isolated sources or small groups of sources which
create a number of small H~II regions. The low-density regions remain largely 
neutral during most of the evolution, since they are devoid of sources.  

The case when only the large (and unsuppressible) sources are present 
(case f2000\_406; Figure~\ref{images}, left) has been discussed in some detail
in Paper I. When smaller sources are also present, the resulting geometry
depends on the radiative feedback upon them. When the low-mass sources are not 
suppressed, the basic morphology is similar to that 
when only larger sources were present (Figure~\ref{images}), with some 
additional small-scale features due to the now-resolved small-mass sources. 
However, when the small-source suppression is included, the morphology changes, 
even with the same source efficiencies (Figure~\ref{images2}). There is more 
small-scale structure present, although the large, locally-percolated regions 
are of similar sizes to the ionized regions observed in the large-box 
simulations, indicating that by this time reionization is already dominated by
the large-mass sources. In the presence of the small-mass, suppressible
sources, there are also small, "relic" H II regions occasionally, i.e. regions
which were ionized earlier but are now recombining because some 
(or all) of the sources which made them originally were suppressed (these
relic H~II regions are seen as darker spots, mostly around the edges of the
large ionized bubbles). Generally, these relic H~II regions are short-lived,
however, 
being quickly overrun by the I-fronts from neighbouring ionized bubbles, and 
thus do not have strong effects on the evolution.      

A direct comparison of the morphology of reionization in the simulations
presented here with those by other methods, such as reported by
\citep{2003MNRAS.343.1101C} or \citep{2006astro.ph..4177Z} is difficult, since
some or all of the properties which determined their outcome, including box
size, resolution, source efficiencies and cosmological parameters, are
generally different. Of these, moreover, only the simulations presented here
were able to resolve the low-mass source halos subject to suppression and took
this suppression into account. We shall present a more detailed discussion of
the characteristic scales and topology of reionization based upon our
simulation results in a forthcoming paper, along with further comparisons. The
purpose of the current paper, however, is to demonstrate the differences made
by improving the resolution to take account of Jeans-mass filtering, as
described above.

\section{Conclusions} 

We have studied the effects of small-mass sources on the progress and duration
of reionization based on a large suite of detailed radiative transfer
simulations. We found that these small-mass sources play an important role
during reionization. In their presence reionization starts much earlier, by
$\Delta z\sim10$ in WMAP1 cosmology (at $z\sim30$ rather than at $z\sim20$), 
and by $\Delta z\sim6$ in WMAP3 cosmology  (at $z\sim22$ rather than at
$z\sim16$). They also supply most or all of the ionizing emissivity at early 
times, when the larger galaxies are still exceedingly rare. 

However, the same low-mass sources are also a subject to suppression in the
ionized regions due to Jeans-mass filtering. This low-mass source suppression
decreases the global ionizing photon emissivity by factors of up to 100,
depending on the photon production efficiency of the small-mass sources.
As a consequence, the Jeans-mass filtering of small-mass sources effects 
the global reionization history profoundly, delaying overlap by $\Delta
z\sim3$, and decreasing the integrated electron-scattering optical depth 
$\tau_{\rm es}$ by $\sim0.025$ compared to the case without suppression. 
The evolving gas clumping at small (sub-grid) scales extends the reionization 
significantly, by $\Delta z\sim1-1.5$, but has only a modest effect on the 
corresponding electron scattering optical depth, decreasing it by
$\sim0.013-0.014$.  
 
Furthermore, the reionization process is strongly self-regulated - the more
efficient the small-mass sources are, the larger the fraction of them that are
suppressed. This results in a corresponding decrease in the number of ionizing
photons emitted, partially canceling the effects of the high efficiency.
However, the later stages of reionization are completely dominated by the
larger-mass sources which do not get suppressed. Their numbers rise
exponentially as they become more common, while, in contrast, the low-mass
halos are strongly suppressed at late times. As a result, the overlap epoch is
not very sensitive to the properties of the low-mass sources, but instead is
largely determined by the numbers and efficiencies of the high-mass
sources. As a consequence, for a given overlap redshift there is a significant
degeneracy between the small-mass source's efficiency and the $\tau_{\rm es}$ 
- by changing the efficiency of the small-mass sources we can easily increase 
or decrease the global integrated optical depth without changing the overlap
redshift.  

In short, the end of reionization is dictated
by the higher-mass haloes which are not subject to Jeans-mass filtering,
but the addition of small-mass, suppressible haloes serves to extend
the reionization epoch to earlier times, thereby boosting $\tau_{\rm es}$.
This naturally explains why WMAP CMB polarization measurements indicate
a large $\tau_{\rm es}$ and early beginning for reionization,  $z > 11$
for WMAP3, while the absorption
spectra for quasars at $z\gsim6$ indicate that the overlap epoch was at
$z < 7$.  Previous simulations failed to see this because they failed to
include the small-mass halo sources with Jeans-mass filtering in a
simulation volume big enough to sample the global reionization history
fairly.

The combined effects of low-mass source suppression in the ionized regions, 
decreasing ionizing source efficiency and increasing gas clumping over time 
can cause
reionization to ``linger'' at a given ionization fraction for some time, 
before eventually reaching overlap. However, none of our models show the
double reionization predicted by \citet{2003ApJ...586..693W} and 
\citet{2003ApJ...591...12C}. What prevents double reionization from 
occurring is the large spread in the reionization histories from one 
region to another, combined with the strong source clustering and the 
self-regulation. Double reionization scenarios require that the 
highly-efficient Pop. III sources emit enough photons to reionize the
universe at high redshift, but then die or become less efficient, 
causing the universe to partially recombine and be finally reionized 
later by the rise of Pop. II sources. However, these highly-efficient
sources form at very different times in different places, and thus the 
transition from high-efficiency to low-efficiency sources is spread 
over time. These early sources also get suppressed quite efficiently, 
thus never get to large enough total emissivity at any given time, so 
as to reionize the whole universe. These self-regulation mechanisms should 
hold under quite generic circumstances, beyond our specific realizations. 
Some of these were previously discussed by \citet{2005ApJ...634....1F} 
within a simplified analytical framework, which did not include some important 
effects like the strongly increased Jeans-mass filtering due to source 
clustering. Nevertheless, they reached conclusions similar to ours, 
namely that double reionization is not physically plausible.

%The reason for this is that all the effects above evolve gradually on 
%the mean, occuring at very different times in different places, rather 
%than suddenly and throughout the universe,
%as these studies assumed. In particular, in our models with source efficiency
%changing from initially high to low, it does so inhomogeneously, at different
%times in different places, and hence the mean effective emissivity drops
%gradually. Based on this, we can conclude that double reionization models are
%rather unphysical. 

Reionization proceeds in an inside-out fashion, whereby the high-density
peaks become ionized first and voids last, thereby confirming the results of
\citet{2006MNRAS.369.1625I}. The mean overdensity of the ionized regions is
always greater than one, and up to 4 at early times. This trend is even
more pronounced in the smaller-box, high-resolution simulations, where the 
density fluctuations are resolved better. In this case also the shapes of
the H~II regions are more non-spherical during their early evolution. Since at
early times both the ionized density field and the source distribution are
strongly biased relative to the underlying density field, it is crucial to
perform detailed numerical simulations in order to obtain the correct
geometry, size and spatial distributions of the ionized bubbles. Most current 
semi-analytical models of reionization ignore the effects of halo bias on the 
Jeans-mass filtering, which leads to a large underestimate of the low-mass 
source suppression. The only exception is the very recent work of
\citet{2006ApJ...649..570K} which presented a model with an approximate, 
spherically-averaged source bias model. While such models cannot give the full
H~II region geometry, it would be interesting to compare their results to full 
simulations in order to see to what extend their statistical results are 
representing the bias effects reliably.

The overlap by redshift $z\sim6-7$ indicated by current observations is
easily achieved by stellar sources with either a Salpeter or a slightly 
top-heavy IMF. There is, therefore, no need for additional, more ``exotic'' 
sources of radiation (e.g. decaying particles) in order to satisfy both the 
electron
scattering optical depth constraint from WMAP and the end-of-reionization
constraints coming from the SDSS QSO's and high-redshift Ly-$\alpha$ surveys,
as they currently stand. 

In WMAP3 cosmology the formation of cosmic structures, and thus the epoch of
reionization, are delayed compared to WMAP1 cosmology, due to a combination of
the lower normalization of the power spectrum of density fluctuations, tilt,
lower matter density and slightly higher Hubble constant. This delay is
such that it almost exactly compensates for the lower value of $\tau_{\rm es}$
found by WMAP3, given ionizing sources with the same efficiency. This was
predicted analytically by \citet{2006ApJ...644L.101A} and confirmed by our
current simulations. 

Our results show that our large-box simulations which include only the
large-mass, unsuppressible sources, nevertheless predict correctly the 
overlap epoch and the large-scale features of reionization. This feature is a
consequence of the strong self-regulation of reionization, due to which by the
time H~II regions grow large (roughly when $x_v\gsim0.1$) the vast majority of
the low-mass sources are already suppressed and the reionization process is
driven by the large-mass sources. However, any simulations which do not
include the low-mass sources would underestimate the ionized fraction,
particularly at early times, and the small-scale features of reionization,
both of which are influenced by the low-mass sources. The total integrated
electron-scattering optical depth is thus also underestimated. Semi-analytical
reionization models which do not include the self-regulation properly would
similarly underestimate the mean ionized fraction history and the
corresponding optical depth. On the other hand, the large-scale features of
reionization are not very sensitive to the low-mass sources and their
efficiencies, but depend significantly on the large-scale features of the
density field, the clustering of the ionizing sources and level of clumpiness
of the gas, all of which are most reliably modelled through large-scale
reionization simulations. 

\section*{Acknowledgments} 
This work was partially supported by NASA Astrophysical Theory Program grants
NAG5-10825 and NNG04G177G to PRS.

\appendix
\section{A Simple toy model for reionization with self-regulation and 
  halo bias}  

\begin{figure}
\includegraphics[width=3.2in]{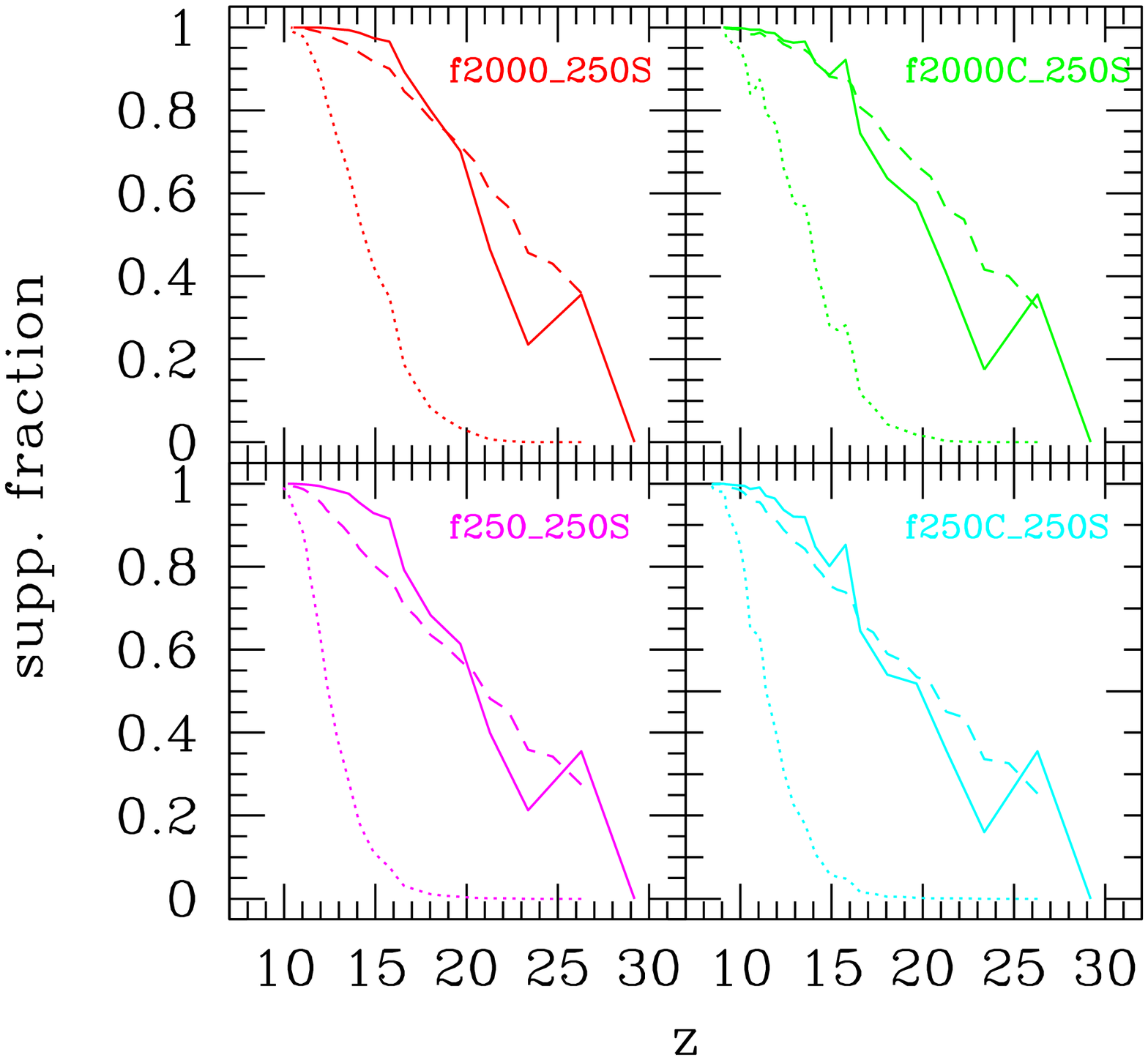}
\caption{The evolution of the fraction of low-mass ionizing sources which are 
Jeans-mass suppressed based on our simulations (solid lines, as labelled), 
a simple model, where the suppressed fraction is simply the mean
volume-weighted ionized fraction (dotted lines), and an empirical model 
where this fraction is $x_v^{0.1}$, to account for source bias (dashed lines).  
\label{supp_frac_35Mpc_wmap1}}
\end{figure}

In order to gain some further insight into the self-regulation of reionization
and its dependence on the assumed parameters, we construct the following simple
toy model. The global average rate of change of the ionized volume fraction
with time, $dx_v/dt$, is determined by the rate of emission of ionizing
photons and the rate of recombinations (per atom in the universe). The
ionizing source model we use in our radiative transfer simulations assumes
that their emissivity is proportional to the collapsed gas fraction in halos, 
$f_{\rm coll}$, with a given efficiency, $f_\gamma$ (see
\S~\ref{eff_sect}). In this, we distinguish the sources whose formation 
is suppressed inside the H~II regions by Jeans-mass filtering from the
high-mass sources, whose formation is not suppressed. In this framework the
rates of 
emission of ionizing photons from the small-mass and large-mass sources are 
given by $f_1\equiv(f_\gamma)_{\rm small}f_{\rm coll,small}/\Delta t$ and 
$f_2\equiv(f_\gamma)_{\rm large}f_{\rm coll,large}/\Delta t$, respectively,
where $\Delta t$ is the time over which these photons are emitted.
The former rate should also be modulated by the ionized fraction, since small
sources are suppressed in ionized regions. In the simplest approximation, one
can assume that the suppressed fraction of small-mass sources is simply
proportional to the ionized volume fraction, $x_v$, i.e. the emitting
(unsuppressed) fraction is given by $f_1(1-x_v)$. Such a model would be exact 
if the small-mass sources were randomly distributed in space. In practice, however,
the halos that host the sources are generally clustered together, thus the
suppression of small-mass sources should be stronger than a random distribution
would give, as these sources cluster around the high-density peaks, which are
ionized earlier in the inside-out reionization yielded by
simulations. Therefore, in an effort to crudely model the effects of this
source bias, we also consider a second case, where we modulate the suppression
by a factor of $b_1(t)=(1-x_v^n)/(1-x_v)$, where we set $n=0.1$ ($n=1$
corresponds to the random distribution above). This power of 0.1 was
determined empirically, by roughly matching the actual Jeans-mass suppressed 
fraction from our simulations to $1-x_v^n$. In
Figure~\ref{supp_frac_35Mpc_wmap1} we plot both models against
the actual data from the simulations (for WMAP1 parameters). We note that this 
simple model is not
intended to match the simulation results perfectly, but rather to show how the
mean reionization histories vary under different assumptions. Thus, even
though it does not follow the Jeans-suppressed source fraction exactly, it  
it is appropriate for our purposes here, as it reflects the main evolutionary 
trends seen in our simulations. 

Finally, the rate of recombinations per unit time per atom is given simply by
the inverse of the recombination time, $f_3\equiv t_{\rm
  rec}^{-1}=C(z)n_H\alpha_B$, where $C(z)$ is the (evolving) clumping factor,
$\alpha_B$ is the Case B recombination rate of hydrogen, and we assume full
ionization in the H~II regions. Combining these factors, we can write the
equation for the evolution of the volume-weighted ionized fraction $x_v$ as:
\be 
\frac{dx_v}{dt}=f_1(t)(1-x_v^n)+f_2(t)-f_3(t)x_v,
\label{model_equ1}
\ee

Based on our simulation data (in WMAP1 cosmology), the functions $f_1$ and
$f_2$ as function of redshift are well-fit by the following expressions:
\ba
\!\!f_1\!\!\!\!\!\!&=&\!\!\!\!\!\!0.335\left(\frac{f_{\gamma,\rm
      small}}{250}\right)\exp\left(0.227z-0.02546z^2\right)\rm Myr^{-1}\\
\!\!f_2\!\!\!\!\!\!&=&\!\!\!\!\!\!4.6312\left(\frac{f_{\gamma,\rm
      large}}{250}\right)\exp\left(-0.107z-0.02463z^2\right)\rm Myr^{-1}, 
\ea
while $f_3$ is given by
\be
f_3=\frac{(1+z)^3C(z)}{6.655\times10^5}\rm Myr^{-1}.
\ee
Finally, in order to use these expressions directly, we need to change the
independent variable in equation~\ref{model_equ1} from time to redshift, using
\be
\frac{dy}{dt}=\frac{dy}{dz}\frac{dz}{dt}
     =-\frac{dy}{dz}H_0(1+z)[\Omega_0(1+z)^3+\Omega_\Lambda]^{1/2}, 
\ee
as appropriate for flat cosmology. The results given by this model are plotted
in Figure~\ref{xv_model_fig} and discussed in the main text. 
\end{document}